\documentclass[twocolumn,aps,pra,superscriptaddress,preprintnumbers,bibnotes]{revtex4-2}

\usepackage[T1]{fontenc}

\usepackage{natbib}
\usepackage{graphicx}
\usepackage{bm}
\usepackage{color}
\usepackage{amsmath}
\usepackage{amssymb}
\usepackage{gensymb} % for \degree
\usepackage{physics}
\usepackage{tabularx}
\usepackage{hyperref}
\usepackage{tikz}
\usepackage{graphicx} % Required for \resizebox

\usepackage{changes}
\definechangesauthor[color=teal]{GP}
\definechangesauthor[color=magenta]{YK}
\definechangesauthor[color=blue]{SG}
\definechangesauthor[color=red]{KK}
\newcommand{\CUAaddress}{Harvard-MIT Center for Ultracold Atoms, Cambridge, Massachusetts 02138, USA}
\newcommand{\HarvardPhysicsAddress}{Department of Physics, Harvard University, Cambridge, Massachusetts 02138, USA}
\newcommand{\HarvardChemistryaddress}{Department of Chemistry and Chemical Biology, Harvard University, Cambridge, Massachusetts 02138, USA}

\DeclareMathOperator*{\sumint}{%
  \mathchoice%
  {\ooalign{$\displaystyle\sum$\cr\hidewidth$\displaystyle\int$\hidewidth\cr}}
  {\ooalign{$\textstyle\sum$\cr\hidewidth$\textstyle\int$\hidewidth\cr}}
  {\ooalign{$\scriptstyle\sum$\cr\hidewidth$\scriptstyle\int$\hidewidth\cr}}
  {\ooalign{$\scriptscriptstyle\sum$\cr\hidewidth$\scriptscriptstyle\int$\hidewidth\cr}}
}

\makeatletter
\newcommand\blfootnote[1]{%
  \begingroup
  \renewcommand\thefootnote{}% Clear the symbol
  \footnotetext{#1}% Use footnotetext to avoid the citation counter
  \makeatletter\addtocounter{footnote}{-1}\makeatother % Ensure counter stability
  \endgroup
}
\makeatother

\begin{document}

\title{Absence of Far-Detuned Attractive Optical Traps for Alkali Rydberg Atoms}

\author{Gabriel E. Patenotte}
\email{gpatenotte@g.harvard.edu; contributed equally.}
% \thanks{G.E.P. and Y.K. contributed equally to this work. Please direct correspondence to gpatenotte@g.harvard.edu.}
\affiliation{\HarvardPhysicsAddress}
\affiliation{\HarvardChemistryaddress} 
\affiliation{\CUAaddress}

\author{Youngshin Kim}
\email{youngshinkim@fas.harvard.edu; contributed equally.}
% \thanks{G.E.P. and Y.K. contributed equally to this work. Please direct correspondence to gpatenotte@g.harvard.edu.}
\affiliation{\HarvardChemistryaddress} 
\affiliation{\CUAaddress}

\author{Samuel Gebretsadkan}
\affiliation{\HarvardPhysicsAddress}
\affiliation{\HarvardChemistryaddress} 
\affiliation{\CUAaddress}

\author{Kang-Kuen Ni}
\affiliation{\HarvardChemistryaddress} 
\affiliation{\HarvardPhysicsAddress}
\affiliation{\CUAaddress}

\date{\today}
\blfootnote{$^{*,\dagger}$\,These authors contributed equally to this work.}
\begin{abstract}
Neutral-atom quantum simulation is susceptible to entanglement between the atom’s internal electronic state and its center-of-mass position. In many alkali Rydberg platforms, the `spin-motion coupling' is exacerbated by the free expansion required to avoid ponderomotive anti-trapping from optical fields. A recent proposal \cite{Bhowmik_VectorTrap_2025} claims sufficiently excited Rydberg states could be trapped in a monochromatic, far-detuned, circularly polarized optical field by harnessing a large vector polarizability. We disprove the proposal through analytic calculation and measurement of the vector polarizability of the $54S$, $54P$, and $53D$ orbitals of Cesium. Regarding the optical angular frequency $\omega$, we analytically derive that the scalar, vector, and tensor polarizabilities scale as $\omega^{-2}$, $\omega^{-3}$, and $\omega^{-4}$, as opposed to the proposed scaling of $\omega^{-2}$, $\omega^{-1}$, and $\omega^{-2}$.  We refine the sum-over-states expression for vector and tensor polarizability to be numerically stable and predict negligible vector and tensor polarizabilities far detuned from resonances, in agreement with our measurements. However, we find vector polarizability can enhance the recent proposal for near-detuned attractive trapping \cite{Jansohn_Kuzmich_MagicWavelengthRydTrap_2025}. Furthermore, we evaluate the breakdown of the electric-dipole approximation and derive no effect stronger than ponderomotive repulsion. We conclude that an attractive, monochromatic, far-detuned optical trap for alkali Rydberg states is not possible, regardless of the beam geometry.

\end{abstract}

\maketitle

\section{Introduction}
\label{sec:Intro}
Long-range electromagnetic interactions are used with great success to generate entanglement. Resonant dipole-dipole interactions and van der Waals interactions have been observed in magnetic atoms \cite{Stuhler2005}, polar molecules \cite{Yan2013}, and Rydberg atoms \cite{Ravets_Browaeys_C3_2014,Beguin_Browaeys_C6_2013}. These dipolar systems have been used to realize quantum gates \cite{isenhower_demonstration_2010,Levine_Lukin_LPGate_2019,Picard_Ni_iSWAP_2025}, improve optical clocks \cite{Eckner_Kaufman_SpinSqueezingRydbergClock_2023}, and generate exotic phases of matter such as spin-liquids \cite{semeghini2021probing}. Many of these advances have been demonstrated with alkali Rydberg atoms, leveraging their large dipole moments, theoretical simplicity, and technically favorable wavelengths.

%Start of new paragraph 2
These applications encode information within the atom's internal states, and are therefore limited by the spatial dependence of the electromagnetic interaction which entangles internal and center-of-mass degrees of freedom. The use of freely expanding Rydberg atoms in many alkali Rydberg experiments exacerbates this `spin-motion coupling' \cite{Emperauger_BenchmarkingDirectIndirect_2025}. This configuration is often chosen for the convenience of using a Gaussian optical tweezer, in which the ground and Rydberg states are generally attracted and repelled, respectively.
%End of new paragraph 2

% %Start of old paragraph 2
% However, the inhomogeneity of the electromagnetic interaction across the particle's wavefunction results in entanglement between internal states and center-of-mass position, limiting applications that encode information in internal states. The use of freely expanding atoms to study dipolar interactions in Rydberg experiments exacerbates spin-motion entanglement \cite{Emperauger_BenchmarkingDirectIndirect_2025}. This configuration is often chosen for the convenience of using a Gaussian optical beam, in which the ground and Rydberg states are generally attracted and repelled, respectively.
% %End of old paragraph 2

All constituent charges of an atom oscillate in an optical field, which increases their kinetic energy inversely with their mass and repels them from intensity. The repulsion can be countered if the optical field couples to internal resonances, creating attractive instantaneous eigenstates whose dipoles are oriented with the optical field. However, alkali Rydberg atoms are repelled due to a lack of strong resonances near optical frequencies. States of the valence electron that differ by optical energies also differ significantly in size, and therefore have transition multipole moments that are suppressed by the Franck-Condon principle.

Significant effort has enabled trapping of alkali Rydberg atoms despite this repulsion. Alkali Rydberg atoms have been confined in the intensity minima of optical lattices \cite{Anderson_TrappedRydbergLattice_2011} and bottle-beam optical tweezers \cite{Li_Saffman_CrossedVortexBoB_2012}, or proposed to be trapped around, but not within, an intensity maximum \cite{Topcu_IntensityLandscape_2013,Cortinas_ThreadingAtom_2024,Rivera_2026_RydTrap}. Nevertheless, retaining the reconfigurability, scalability, and proximity of a standard optical tweezer platform remains challenging with repulsive schemes.  Attractive traps have been proposed by detuning close to a weak optical resonance \cite{Jansohn_Kuzmich_MagicWavelengthRydTrap_2025}; however, the admixture of a compact excited orbital enhances scattering, and the trap is state-specific. In contrast, divalent alkaline earth Rydberg atoms offer state-insensitive attractive trapping via the strong optical transitions of the non-Rydberg core electron \cite{Wilson_2022_Thompson_TrappedRydbergYb}.

We investigate a proposal to trap alkali Rydberg atoms \cite{Bhowmik_VectorTrap_2025} using vector polarizability. A circularly polarized tweezer produces a fictitious magnetic field proportional to the optical intensity, such that states whose permanent magnetic moments are oriented with the field experience attraction towards intensity. The authors performed a sum over bound states and found the vector polarizability for sufficiently excited Rydberg atoms to exceed the ponderomotive effect due to more favorable scaling with the optical angular frequency ($\omega^{-1}$ vs $\omega^{-2}$).

In section \ref{sec:Definition} we define polarizability. In sections \ref{sec:Measurement} and \ref{sec:Result} we measure the polarizability of the fine structure levels within the $54S$, $54P$, and $53D$ orbitals of Cesium and report negligible vector and tensor polarizabilities, contrary to \cite{Bhowmik_VectorTrap_2025}. In sections \ref{sec:TheoryOverview} to \ref{sec:TheoryBeyondDipole} we analytically prove vector and tensor polarizabilities are negligible for far-detuned optical fields, with and without assuming the electric dipole approximation.  Lastly, in section \ref{sec:NearDetuned} we calculate the scattering for a near detuned attractive trap \cite{Jansohn_Kuzmich_MagicWavelengthRydTrap_2025} and find reduction using circular polarization.  % To validate the proposal, in Sec. \ref{sec:Measurement} we measure the vector polarizability of Zeeman sublevels in the 54S, 54P, and 53D orbitals of Cesium using both linearly and circularly polarized optical tweezers. Contrary to the theoretical prediction, we observe a negligible vector polarizability. We resolve this discrepancy in Sec. \ref{sec:TheoryNotBeyondDipole}. by developing a theoretical framework demonstrating that the leading-order polarizability of Rydberg atoms remains ponderomotive, even when accounting for the breakdown of the electric-dipole approximation. We demonstrate that the leading order vector and tensor polarizability terms scale as $\omega^{-3}$ and $\omega^{-4}$, and are due to the the spin-orbit interaction and the anharmonicity of the Coulombic potential. In addition, we provide accurate expressions for the vector and tensor polarizabilities in a sum-over-states calculation.
\section{Definition}
\label{sec:Definition}

In the absence of external fields, the valence electron of an alkali Rydberg atom is well described in the fine structure basis with the notation $\ket{nLSJM}_{\hat{z}}$. Here $n$ is the principal quantum number, $\hbar\sqrt{L(L+1)}$ and $\hbar\sqrt{S(S+1)}$ are the magnitudes of orbital ($\vec{L}$) and spin ($\vec{S}$) angular momenta, $\hbar\sqrt{J(J+1)}$ is the magnitude  of the total angular momentum ($\vec{J}=\vec{L}+\vec{S}$), and $\hbar M$ is the component of $\vec{J}$ along the axis $\hat{z}$. The ionic core, consisting of the nucleus and inner-shell electrons, has negligible spatial overlap with the valence electron and interacts weakly with external fields, allowing it to be treated as a spectator.

We first address the validity of the electric-dipole approximation for a Rydberg atom in a weak, off-resonant electric field from an optical tweezer of wavelength $\lambda$. The electric-dipole approximation is valid when $kr_\text{rms}\ll 1$ where $k=2\pi/\lambda$ and $r_\text{rms}$ is the root-mean-square radius of the valence electron:
\begin{align}
   r_\text{rms}=\sqrt{\langle r^2\rangle_{nl}}=n\sqrt{\frac{5n^2+1-3l(l+1)}{2}}a_0
\end{align}
This radius scales from $\sim 0.1$ nm for ground-state atoms to $\sim 200$ nm for the Rydberg states measured in this work. In comparison, the longitudinal phase and transverse amplitude of the 1064 nm optical tweezer vary on the scale of the wavelength with a wavenumber of $k=(170 \text{ nm})^{-1}$. Consequently, the breakdown of the electric-dipole approximation is a non-negligible effect for Rydberg atoms. We introduce the atom-field interaction within the electric-dipole approximation here and in Sec. \ref{sec:NotBeyondDipole}, and analyze contributions beyond the dipole approximation in Sec. \ref{sec:TheoryBeyondDipole}.
\begin{figure}[htbp]
    \centering
    % This resizes the tikzpicture to fit the column width
    \resizebox{0.8\columnwidth}{!}{\input{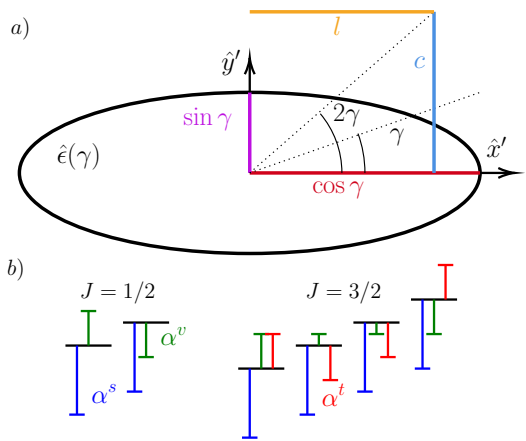}}
    \caption{(a) The polarization ellipse for an electric field with ellipticity $\gamma$, linearity $l=\cos(2\gamma)$, and circularity $c=\sin(2\gamma)$. The electric field is $\vec{E}(t)=\Re(\mathcal{E} \hat{\epsilon}e^{i(kz-\omega t)})$ with the ellipse showing the directions that the electric field points over a period of oscillation. Both the direction of field propagation and the effective magnetic field experienced by the atom are into the page. (b) A schematic for how the scalar, vector, and tensor polarizabilities diagonally perturb the sublevels of a fine structure level of $J=\{1/2,3/2\}$ that are quantized by a DC magnetic field. The sublevels are arranged in order of increasing $m_J$, and the arrows depict how the scalar, vector, and tensor polarizabilities apply a common-mode, antisymmetric, and symmetric change to the energy of the sublevels.}
    \label{fig:conceptual}
\end{figure}

A uniform, monochromatic electromagnetic field $\vec{E}(t)=\Re(\vec{\mathcal{E}}e^{-i\omega t})$ of angular frequency $\omega$ may be defined in a new frame where $\hat{z}'$ is the direction of propagation and the polarization is restricted to the $\hat{x}'$-$\hat{y}'$ plane. The field amplitude is described by a magnitude $\mathcal{E}\equiv|\vec{\mathcal{E}}|$ and ellipticity $\gamma\in[-\frac{\pi}{4},\frac{\pi}{4}]$ such that $\vec{\mathcal{E}}=\mathcal{E}\left(\cos\gamma\,\hat{x}'+i\sin\gamma\,\hat{y}'\right)$. The circularity ($c$) and linearity ($l$) of this light are related to the ellipticity by $c=\sin(2\gamma)\in[-1,1]$ and  $l=\cos(2\gamma)\in[0,1]$, respectively. As defined, $c=1$ corresponds to light that propagates along $\hat{z}$ with its polarization rotating with the right-hand rule, and it drives an atom's $\hat{\sigma}_+(\hat{\sigma}_-)$ transition when the internal state is excited (de-excited). The polarization ellipse of the field is shown in Fig. \ref{fig:conceptual}(a). 

This electric field perturbatively couples the atomic states, shifting energies and mixing states within each level. As derived in Appendix \ref{appendix:PerturbationTheory}, the second order perturbation for a level $\ket{nLSJ}$ is 

\begin{align}
    \delta H^{(2)}=&-\frac{\mathcal{E}^2}{4}\left(\alpha^s+c\frac{\hat{J}_{z}}{2J}\alpha^v-\frac{3(\hat{J}_{\hat{z}}^2-l\frac{\hat{J}_{-}^2+\hat{J}_{+}^2}{2})-\hat{J}^2}{2J(2J-1)}\alpha^t\right)
    \label{eq:svtDef}
\end{align}

where we choose the propagation frame to quantize $\hat{J}_{z}$ and $\hat{J}_\pm$. The perturbation is parameterized by the scalar ($\alpha^s$), vector ($\alpha^v$), and tensor ($\alpha^t$) polarizabilities and is depicted in Fig. \ref{fig:conceptual}(b). The polarizabilities depend on the quantum numbers $nLSJ$ and the optical angular frequency. The vector and tensor terms are nonzero only if $J\geq 1/2$ and $J\geq 1$, respectively.

We focus on the vector polarizability, motivated by the prediction in Ref. \cite{Bhowmik_VectorTrap_2025} that $|\alpha^v|$ may exceed $|\alpha^s|$. To understand the signature of $\alpha^v$, we consider the effect of a static magnetic field $\vec{B}_\text{DC}$. It interacts with the permanent magnetic moment of the fine structure level to produce a first-order Zeeman shift: 
\begin{align}
    \delta H^{(1)}&=-\vec{\mu}_J\cdot\vec{B}_\text{DC}\\
    \vec{\mu}_J&=-\frac{g_J\mu_B}{\hbar}\vec{J}
\end{align}
By inspection of the Hamiltonian $\delta H^{(2)}$, the term proportional to $\alpha^v$ mimics this magnetic interaction. Defining the vector light shift in terms of a `fictitious' magnetic field $\vec{B}_\text{fict}$ parallel to the propagation vector $\vec{c}=c\hat{z}$, we identify the vector component of the perturbation as
\begin{align}
    \delta H^{(2)}\bigr|_\text{vector} &=-\vec{\mu}_J\cdot\vec{B}_\text{fict}\\
    \vec{B}_\text{fict}&=\frac{\hbar\alpha^v}{g_J\mu_B}\frac{|\vec{\mathcal{E}}|^2}{4}\frac{\vec{c}}{2J}
\label{eq:Bfict}
\end{align}
The effective perturbation to the fine structure level is thus $\delta H=\delta H^{(1)}+\delta H^{(2)}=-\vec{\mu}_J\cdot\vec{B}_\text{eff}$ where $\vec{B}_\text{eff}\equiv\vec{B}_\text{ext}+\vec{B}_\text{fict}$. In the following section, we infer $\alpha^v$ by performing spectroscopy on the Zeeman sublevels of several Rydberg fine structure levels, observing the modification of the splitting caused by the fictitious field $\vec{B}_\text{fict}$.

In analogy, the tensor polarizability produces a fictitious electric field gradient that interacts with the permanent electric quadrupole moment of the fine structure level. Instead, however, we find the diagonal contribution of the tensor polarizability when the propagation axis is at an angle $\theta$ and $\phi$ from the DC magnetic field to be
\begin{align}
    \delta E^{(2)}_\text{ac}\big|_\text{t}&=\frac{\mathcal{E}^2}{4}\alpha^t d(\theta,\phi)\frac{3M^2-J(J+1)}{2J(2J-1)}\\
    d(\theta,\phi)&=\frac{3\cos^2(\theta)-1}{2}-l\frac{3}{2}\cos(2\phi)\sin^2(\theta)\label{eq:dDef}
\end{align}
The angular term $d\in [-2,1]$ evaluates to $d(0,0)=1$ for this experiment's geometry.
\section{Measurement}\label{sec:Measurement}
\begin{figure}[htbp]
    \centering
    % This resizes the tikzpicture to fit the column width
    % \resizebox{\columnwidth}{!}{\input{Figures/figureApparatus}}
    \includegraphics[width=\columnwidth]{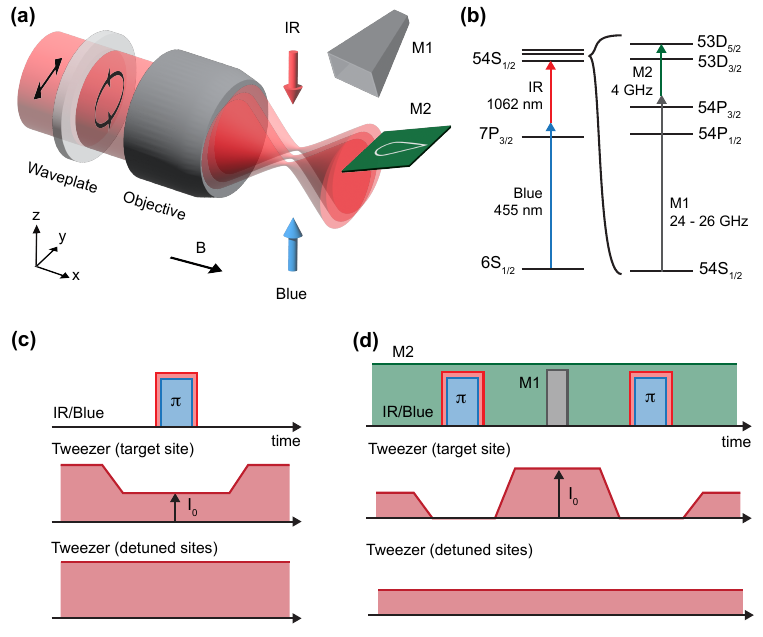}
    \caption{(a) Experimental apparatus. Atoms are trapped in an optical tweezer array, whose ellipticity is controlled by a waveplate on a motorized rotation mount. (b) Energy levels relevant to this work. Rydberg state $54S_{1/2}$ is reached by two vertical laser beams (`IR' and `Blue'), and additional states $54P$, $53D$ are addressed by two microwave radiations (`M1' and `M2'). (c) Pulse sequence for probing polarizability of $54S_{1/2}$ states. Power $P$ of tweezer light is varied only for a target site to excite one atom at a time, and this step is repeated for all sites. (d) Pulse sequence for probing polarizability of $54P$ and $53D$ states. IR/Blue $\pi$-pulse is applied to prepare $54S_{1/2}, m_J = 1/2$ at a target site, and another $\pi$-pulse is applied for read-out. The power of tweezer is varied in between, during which microwave M1 is pulsed to measure light shift (M2 is continuously on).}
    \label{fig:apparatus}
\end{figure}
The apparatus used to measure the vector polarizability of the  $54S_{1/2}$, $54P_{1/2}$, and $53D_{3/2}$ fine structure levels has been described previously \cite{zhang_optical_2022,park_extended_2023} and is illustrated in Fig. \ref{fig:apparatus} (a). Individual Cs atoms are loaded into a 1064~nm 1D optical tweezer array from a magneto-optical trap in an evacuated glass cell.  The array is formed by an acousto-optic beam deflector (AOBD) powered by radio-frequencies that are generated by a 625 MHz sampling rate arbitrary waveform generator (AWG, Spectrum Instruments M4i.6622-x8). The amplitudes of each tone are normalized to make the radial trap frequencies of each site to be the same within 1\%, calibrated via parametric heating resonances. The polarization of the tweezer is made linear by a Glan-Taylor polarizer, and the beam then passes through a $0.304\lambda$ waveplate in a motorized rotation mount (Griffin Motion RTS100) that can scan the circularity of the light between $c\in [-0.94,0.94]$ by a rotation of $\pm 45$ degrees. 

All states are labeled in the fine-structure basis with $\ket{nL_J,M_J}\ket{I,M_I}$ notation. Optical pumping prepares the stretched state $\ket{g}\equiv\ket{6S_{1/2},1/2}\ket{7/2,7/2}$, whose quantization is enforced by a $6$ Gauss external magnetic field that can be arbitrarily rotated using three pairs of coils. Polarization gradient cooling (PGC) prepares the atoms at a temperature of $\sim10$~$\mu$K \cite{liu_molecular_2019}. The atom is excited to the Rydberg state $ 54S_{1/2}$ level from $\ket{g}$ using a two-photon transition that is 1 GHz blue detuned of the intermediate $7P_{3/2}$ level, as shown by the energy level diagram in Fig. \ref{fig:apparatus} (b). The 455~nm and 1064~nm used for the two-photon transition are locked to a $\lesssim 100$ kHz linewidth cavity, as described in a prior work \cite{Fang_Critical_2025}. The excitation beams counter-propagate along $\hat{z}$ with approximately balanced Rabi frequencies and oppositely circular polarizations; however the quantization axis during the excitation step is chosen to be $\hat{x}$ such that $\ket{g}$ can couple to both $\ket{S_-}\equiv\ket{54S_{1/2},-1/2}$ and $\ket{S_+}\equiv\ket{54S_{1/2},1/2}$. Both transitions can be coherently populated with a $\pi$ time of $\sim 1$ $\mu$s. Note that the nuclear spin is uncoupled from the fine structure in Rydberg states due to small overlap of the valence electron with the nucleus, so the notation $\ket{I,M_I}=\ket{7/2,7/2}$ can be omitted. Each excitation beam is pulsed using an acousto-optic modulator (AOM). The AOMs are driven by mixing a fixed radio-frequency signal near $180$~MHz with a DC pulse from an arbitrary function generator (Tektronix AFG 31052).

The sequence for measuring the polarizability of $54S_{1/2}$ is described as follows, and illustrated in Fig. \ref{fig:apparatus} (c). After rearrangement \cite{Picard_SiteSelective_2024}, we prepare a nearly defect-free array of 8 Cs atoms spaced by 5~$\mu$m, and take a first fluorescence image sensitive to all hyperfine states in $6S_{1/2}$ to check for the presence of an atom in each tweezer. The atoms are optically pumped in $\ket{g}$ while quantized by a 6 Gauss magnetic field along $\hat{y}$, following which the optical tweezer power is lowered to a minimal depth and the magnetic field is rotated to $\hat{x}$ over 5 ms. The rotation mount is then rotated to a scannable angle in 82 ms, with an additional 68 ms pause for residual oscillations of the rotation angle to settle to within 0.001 degrees. After adiabatically restoring the tweezer powers to their maximum, we diabatically drop the tweezer on the target site to a scannable depth in $0.5$~$\mu$s, while the other sites remain deeply trapped. The excitation beams are pulsed using scannable durations and frequencies. The optical tweezer is turned back on for 100 $\mu$s to eject the Rydberg atom population. The target site is iterated over all 8 sites, sequentially exciting each atom, after which the population in $6S_{1/2}$ is detected via fluorescence imaging. The retention probability of $6S_{1/2}$ is obtained from the atom survival conditioned on its presence in the initial imaging. In this manner, for each chosen tweezer power and polarization, the frequencies of the two resonances to $54S_{1/2}$ are identified at their $\pi$ time, which provides sufficient information to measure the scalar and vector polarizability of the fine structure level.

To identify any systematic error due to the diabatic tweezer ramps or the sequential excitation scheme, we also perform a separate measurement of polarizability $54S_{1/2}$ on two atoms separated by $\sim 20$~$\mu$m. We simultaneously lower the two tweezers adiabatically to a scannable depth in $3$~ms, apply excitation pulse, and increase it back for the Rydberg-atom ejection. This sequence yields similar results for $54S_{1/2}$ polarizability, which we will discuss in the next section.

The polarizability of the $54P_{1/2}$ and $53D_{3/2}$ levels is measured using two additional microwave fields, M1 and M2. The $54S_{1/2}$ to $54P_{1/2}$ transition is addressed using the M1 microwave horn antenna powered by a $\sim$ 23 GHz signal generator (DS Instruments SG30000L) mixed with a DC pulse from an AFG (Keysight 33622A). The $54P_{1/2}$ to $53D_{3/2}$ transition is addressed using the M2 PCB antenna powered by a $\sim$6 GHz signal generator (Valon Technology 5009) mixed with a DC pulse from the other channel of the same AFG. For spectroscopy of  $54P_{1/2}$ the aforementioned sequence is adapted, as shown in Fig. \ref{fig:apparatus} (d). In addition to the 6 Gauss magnetic field from the shim coils, 50 Gauss from a pair of larger coils is applied in the same direction to separate magnetic sublevels, which is ramped over 40~ms before the excitation step. Following the excitation pulse to $\ket{S_+}$, a M1 microwave pulse to $54P_{1/2}$ of scannable duration, frequency and power is applied, following which the population remaining in $\ket{S_+}$ is restored to $\ket{g}$. The ejection and imaging steps then proceed as before. For spectroscopy of $53D_{3/2}$ and $53D_{5/2}$, the single-photon microwave pulse to $54P_{1/2}$ is replaced by a detuned two-photon pulse that resonantly drives $\ket{S_+}$ to $53D_{3/2}$, where the frequency is fixed at $23.7$~GHz while M2 frequency is scanned near $4$~GHz.

\section{Experimental Result}
\label{sec:Result}
\begin{figure}[htbp]
    \centering
    % This resizes the tikzpicture to fit the column width
    % \resizebox{\columnwidth}{!}{\input{Figures/figure_gs_result}}
    \includegraphics[width=\columnwidth]{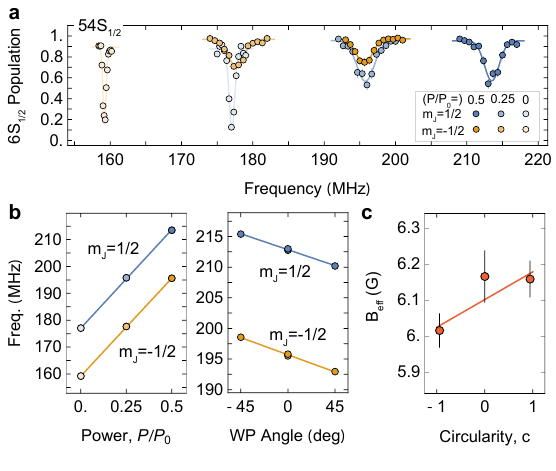}
    \caption{Light shift measurement for $54S_{1/2}$. (a) Excitation spectra of $m_j = \pm1/2$ levels at three different tweezer powers $P/P_0 = 0, 0.25, 0.5$, where $P_0$ is the initial trap depth. The y-axis is the probability of remaining in the ground level  $6S_{1/2}$ after a Rydberg excitation. The x-axis is a scan of the Blue laser's frequency about 657933.21 GHz with the IR laser at 282286.14 GHz. The trap depths of the detuned sites are maintained at $P_0$. (b) (left) Frequency shift vs. tweezer power for the two $m_j$ levels. The curves represent linear fits. (right) Frequency shift vs. waveplate angle at the tweezer power of $P/P_0 = 0.5$. (c) Effective total magnetic field $B_{\text{tot}}$ as a function of circularity, $c$ (see text).}
    \label{fig:gs}
\end{figure}

%1168 from ARC
The measurement results for $54S_{1/2}$ is shown in Fig. \ref{fig:gs}. We drive $\ket{g}$ to $\ket{S_\pm}$ transitions for their respective $\pi$ times (0.7 and 1.5 $\mu$s) at varying optical tweezer powers with linear polarization to calibrate the optical field strength. We expect the shift of resonance frequency to be proportional to the differential scalar polarizabilities between $\ket{g}$ and $\ket{S_\pm}$ as $\delta E = -(\mathcal{E}^2/4) (\alpha_{54S_{1/2}}^s - \alpha_{6S_{1/2}}^s)$. Using the known values in the literature at $\lambda = 1064$~nm for $\alpha_{6S_{1/2}}^s(\omega) = 1163~\mathrm{a.u.}$  and $\alpha_{54S_{1/2}}^s(\omega) = -538~\mathrm{a.u.}$ \cite{ARC} (close to the free electron value $\alpha_{54S_{1/2}}^s(\omega) = -e^2/m \omega^2 = -545~\mathrm{a.u.}$), we estimate the field amplitude to be 
$\mathcal{E} = 2.63 \pm 0.01  $~V/$\mu$m 
%$\mathcal{E} = 2.628 \pm 0.008  $~V/$\mu$m 
from the linear fits of the measurements (independently, we obtain 
$\mathcal{E} = 2.50 \pm 0.01 ~V/\mu$m
%$\mathcal{E} = 2.498 \pm 0.005 ~V/\mu$m
from the two-atom measurement). Due to finite initial temperatures and the expansion of the motional wavefunction, this represents an effective, site-averaged value of the tweezer field experienced by the atoms. We also observe broadening of peaks as the tweezer power increases. We attribute this to site-by-site trap depth inhomogeneity %Our trap frequency normalization does not eliminate this, implying trap-by-trap abberations. 
and the anti-trapping of the Rydberg states; the latter affects $\ket{S_-}$ more due to the longer pulse time.

For vector polarizability, we fix the tweezer power and rotate the waveplate to measure the shift of the $m_J = \pm1/2$ peaks (Fig \ref{fig:gs} b). We fit the splitting of $m_J$ magnetic sublevels to $\delta E = g_J m_J B_{\text{tot}}$ to obtain the effective total magnetic field $B_{\text{tot}} = B_{\text{DC}} + B_{\text{fict}}$ for each circularity $c$ (waveplate angle). From the slope of $B_{\text{tot}}$ as a function of $c$, we can extract the value of vector polarizaiblity $\alpha^v$ using the equation (\ref{eq:Bfict}).
%The energy shift will be proportional to the circularity of light as $\Delta U_{m_J} = -(\mathcal{E}^2/4) (m_J \alpha_{54S}^v - (1/2)\alpha_{6S}^v)c $. The difference $\Delta U_{1/2} - \Delta U_{-1/2}$ is only sensitive to the shifts of the magnetic sublevels $\ket{S_\pm}$. By comparing this to the measured shift with circular polarizations and using the $\mathcal{E}$ value calibrated in the linear tweezers, we estimate the vector polarizability to be $\alpha_{54S}^v = -10 \pm 7$~a.u. 
We estimate the value to be $\alpha_{54S}^v = -10 \pm 5$~a.u. This is surprisingly small compared to the predictions of $\sim 1800$~a.u. from references \cite{ARC,Bhowmik_VectorTrap_2025} . In the following section, we will show that the discrepancy comes from the numerical instability in the sum-over-state calculations for the Rydberg states.

We can also compute the sum of the energy shifts of the two $m_J$ states $\delta E_{1/2} + \delta E_{-1/2}$ to isolate the ground-state contribution, where the excited-state contributions cancel out because $\delta E_{m_J} = -(\mathcal{E}^2/4) (m_J \alpha_{54S}^v - (1/2)\alpha_{6S}^v)c $. We obtain the ground-state vector polarizability to be $\alpha_{6S}^v = -267 \pm 10$~a.u.
The result does not directly agree with the numerical calculation, $\alpha_{6S_{1/2}}^v(\omega) = -200$~a.u \cite{ARC}. As explained in Section \ref{sec:Measurement}, we also conduct a separate measurement on two distant atoms with an adiabatic tweezer ramp sequence, and obtain $\alpha_{6S}^v = -227 \pm 18$~a.u. These results indicate that there could be as large as $\sim 30 \%$ systematic error on our vector polarizability measurements potentially induced by diabatic tweezer ramps and tightly-focused tweezer light, %which does not change the main conclusion of this work.
which may cause atomic wavefunctions to sample different parts of the tweezer. We emphasize that the main conclusion of this work does not change even in the presence of this systematic shift.

\begin{figure}[htbp]
    \centering
    % This resizes the tikzpicture to fit the column width
    %\resizebox{\columnwidth}{!}{\input{Figures/figure_PD_result}}
    \includegraphics[width=\columnwidth]{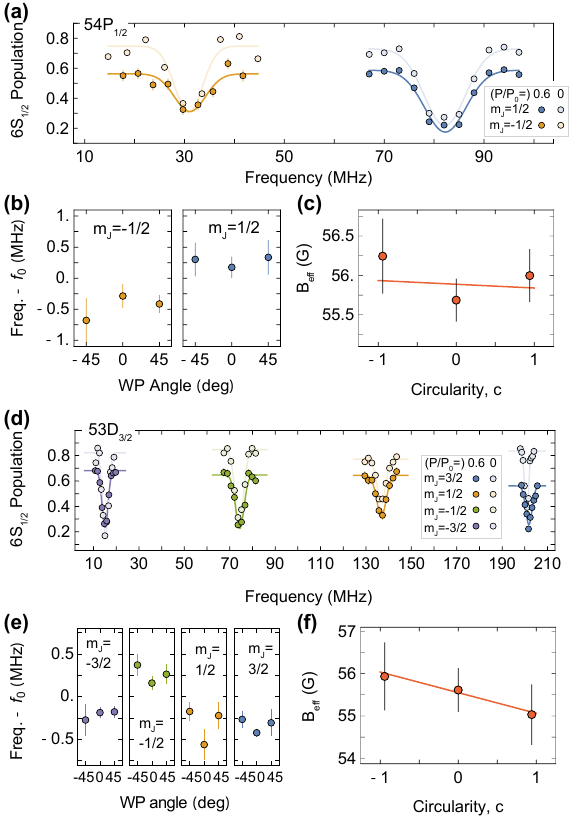}
    \caption{Light shift measurements for $54P_{1/2}$ and $54D_{3/2}$. (a,d) Excitation spectra of magnetic sublevels in $54P_{1/2}$ and $53D_{3/2}$ at two different tweezer powers $P/P_0 = 0, 0.6$. The x-axis represents M1 and M2 microwave frequencies, plotted relative to 23.7 GHz and 3.9 GHz, respectively. (b,e) Frequency vs. waveplate angle for the magnetic sublevels, plotted relative to their respective bare resonances without the tweezers. (c,f) Effective total magnetic field as a function of circularity, $c$. Error bars in (b) and (e) are the standard error of mean of centers from Gaussian fits for least 5 measurements, each comprising several hundred runs.}
    \label{fig:PD}
\end{figure}

The conventional theory also predicts a large dependence of vector polarizability on the orbital angular momentum \cite{Bhowmik_VectorTrap_2025}. Our measurements on $L=1,2$ states are shown in Fig. \ref{fig:PD}. We drive transitions from $\ket{S_+}$ to the magnetic sublevels of $54P_{1/2}$ and $53D_{3/2}$ manifolds roughly at their respective $\pi$ times ($0.2$ - $0.6$ $\mu$s). 
%(Here we denote $\ket{P_\pm} = \ket{54P_{1/2},\pm1/2}$, $\ket{D_\pm} = \ket{53D_{3/2},\pm1/2}$, and $\ket{D_{\pm\pm}} = \ket{53D_{3/2},\pm3/2}$.) 
Because the scalar polarizabilities of Rydberg states are almost identical, there is no significant dependence of resonance frequencies on the optical power in the linear tweezers, as shown in Fig. \ref{fig:PD} (a) and (d). For the vector polarizability measurements, we fix the power and rotate the waveplate (Fig. \ref{fig:PD}).
As with the analysis of $54S_{1/2}$ states, we obtain the effective total magnetic fields $B_{\text{tot}}$ for each circularity $c$, and estimate the vector polarizability from the dependence of $B_{\text{total}}$ on $c$.
% For each magnetic sublevel, the transition frequency is expected to follow $\Delta U_{m_J} = -(\mathcal{E}^2/4) [ (m_J/2J) \alpha_{nL_J}^v - (1/2) \alpha_{54S}^v ]c$. A weighted sum $\sum_{m_J} \mathrm{sgn}(m_J) \Delta U_{m_J} = -(\mathcal{E}^2/4) [(J+1/2)^2/(2J)]\alpha_{nL_J}c$ is only dependent on the upper state vector polarizability. Comparing this to the measured sums in Fig. \ref{fig:PD} (c) and (f), along with the calibrated value of $\mathcal{E}$, 
We estimate that the values for $54P_{1/2}$ and $53D_{3/2}$ are consistent with being close to zero, $\alpha_{54P_{1/2}} = 2 \pm 12$~a.u. and $\alpha_{53D_{3/2}} = 64 \pm 73$~a.u.,
%{We estimate that the vector polarizabilities of $54P_{1/2}$ and $53D_{3/2}$ are close to zero, $\alpha_{54P_{1/2}} = 5 \pm 9$~a.u. and $\alpha_{53D_{3/2}} = 1 \pm 7$~a.u,} 
finding no substantial dependence on $L$.

\begin{figure}[htbp]
    \centering
    % This resizes the tikzpicture to fit the column width
    %\resizebox{\columnwidth}{!}{\input{Figures/figure_PD_result}}
    \includegraphics[width=\columnwidth]{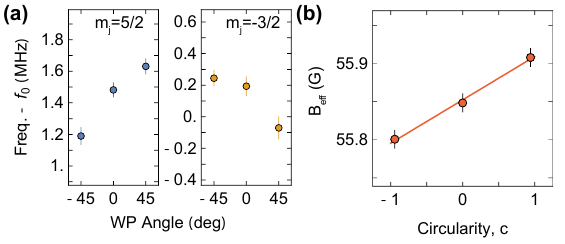}
    \caption{Light shift measurements for $54D_{5/2}$. (a) Frequency vs. waveplate angle for two magnetic sublevels, $m_J = 5/2,-3/2$, at the tweezer power $P/P_0 = 0.75$. The data is plotted relative to the bare resonances without the tweezers. (The microwave powers of M1 and M2 were reduced for $m_J = 5/2$ to increase resolution, resulting in the overall frequency shift of $\sim 1.5$~MHz due to light shift.) (b) Effective total magnetic field as a function of the circularity, $c$.}
    \label{fig:D52}
\end{figure}

In \ref{sec:VectorPol}, we predict a small but measurable value for $\alpha_{53D_{5/2}}^v(\omega) = -14$~a.u, due to the $1064$~nm tweezer light being on a tail of resonance. We measure this via two magnetic sublevels $m_J = 5/2,-3/2$ (Fig. \ref{fig:D52}), and indeed find a small non-zero value, $\alpha^v_{53D_{5/2}} = -15 \pm 2$~a.u. In this measurement, we rotated the magnetic field after the $150$-ms waveplate rotation step (cf. \ref{sec:Measurement}), which reduced the broadening of resonance peaks and led to smaller uncertainties.

The states with $J>1/2$, e.g. $53D_{3/2}$ and $53D_{5/2}$, can also experience tensor shift even in a linear tweezer, according to (\ref{eq:svtDef}). We can estimate the tensor polarizability $\alpha^t$ by measuring the energy shift induced by a linearly-polarized tweezer, and fitting it to $\delta E_{m_J} \propto (-\mathcal{E}^2/4)(a_0 + a_2 m_J^2)$. From the coefficient of the quadratic term, we find that the values are $\alpha^t_{53D_{3/2}} = 9\pm40$ and $\alpha^t_{53D_{5/2}} = -6\pm 11$.

In all analysis of the measured $\alpha^v$ and $\alpha^t$, we estimated uncertainties in two ways. In the first method, we propagate the errors in each measurement $(x_i, y_i)$ via the covariant matrix $C = (J^TWJ)^{-1}$, where $J_{ij} = \partial f(x_i)/\partial a_j$ is the Jacobian, $f(x)$ is the model function, $a_j$ are the fit parameters, and $W_{ii} = 1/\sigma_{i}^2$ are the weights computed from the uncertainties $\sigma_{i}$. The parameter uncertainties are given by the square roots of the diagonal elements, $\sqrt{C_{ii}}$. In the second method, we multiply $C$ by a scale factor $\hat{\sigma} = (n-p)\sum_i (y_i - f(x_i))^2/\sigma_i^2$ where $n$ is the number of measurements and $p$ is the number of parameters. The second method is more appropriate when the residuals between the measurements and the model are larger than the estimated statistical uncertainties due to systematic errors. We report the larger of the uncertainties obtained from the two methods.

\begin{figure}[htbp]
    \centering
    % This resizes the tikzpicture to fit the column width
    %\resizebox{\columnwidth}{!}
    \includegraphics[width=0.9\columnwidth]{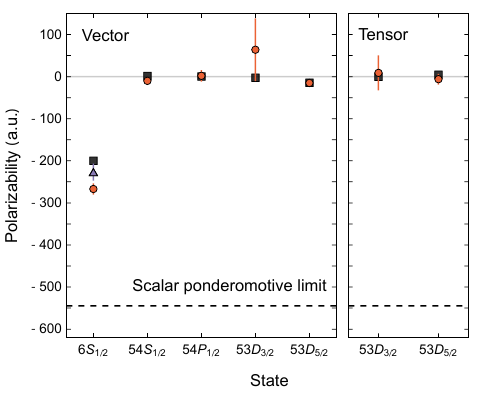}
    \caption{Summary of vector and tensor polarizability measured in this work (red) vs. calculated values from perturbation theory (black) at $1064$~nm. Additional data for $54S_{1/2}$ taken with adiabatic tweezer ramp is shown in purple (see text). Calculations are done with the numerically stable method of section \ref{sec:NotBeyondDipole} using reduced matrix elements from the ARC python package \cite{ARC}. The ponderomotive polarizability is indicated as a dashed line. Error bars represent the standard error from the fits (see text) and systematic error (2\%) in tweezer intensity for different ellipticities. }
    \label{fig:summary}
\end{figure}

A summary of all the vector polarizability measurements as well as the predictions from our refined theory is shown in Fig. \ref{fig:summary}.

\section{Theory Overview}
\label{sec:TheoryOverview}
We find that a far-detuned optical tweezer is unable to break the degeneracy of a fine structure level. This implies the absence of vector and tensor effects, which can be fully explained by the theory developed in Sections \ref{sec:NotBeyondDipole} to \ref{sec:TheoryBeyondDipole}. We first provide a broad overview of the derivation and the main theoretical results. 

An atom in a weak, uniform optical field experiences a second order perturbation due to the electric-dipole interaction $V_\text{E1}=-\vec{d}\cdot \vec{E}(t)$. In (\ref{eq:secondOrderAF}) we show the perturbation may be defined as the contraction of two spherical tensors
\begin{align}
\delta H^{(2)}_\text{ac}&=\sum_{\kappa=0}^2 (-1)^\kappa\alpha^{(\kappa)} \cdot\mathcal{F}^{(\kappa)}\nonumber
\end{align}
which are the atom's polarizability $\alpha^{(\kappa)}$, defined in (\ref{eq:pol_dRd}), and a field tensor $\mathcal{F}^{(\kappa)}$, defined in (\ref{eq:FieldTensor}). 

The polarizability depends on the optical angular frequency $\omega$ and on the transition angular frequency between internal states of the atom, $\Delta_0$, defined in (\ref{eq:transitionAngularFrequencyOperator}). We use an operator Taylor series to expand the polarizability $\alpha^{(\kappa)}$ in powers of  $\Delta_0/\omega$. The series rapidly converges for the dominant microwave transitions of a Rydberg atom, but not for its weak optical transitions. In sections \ref{sec:DecompResolvent} to \ref{sec:ScalarTensorPol} we find the first order term of the series to be analytically zero for $\kappa=1$ and $\kappa=2$, and we prove that the leading order dependence is $\omega^{-3}$ and $\omega^{-4}$, respectively. We return to the unexpanded form of $\alpha^{(\kappa)}$ and subtract away these `zero' term. Our result is an expression for the polarizability that is nominally identical to the one in the literature \cite{Kien2013Polarizability}, but is numerically stable against errors in the matrix elements of the strong microwave transitions between adjacent Rydberg states. 

A comparison between the original and modified expression for the polarizability is clearest in the definition of its reduced matrix element for a fine structure level:
\begin{align}
    &\bra{nLSJ}|\alpha^{(\kappa)}|\ket{nLSJ}\nonumber\\&=e^2\sum_{L'J'}\mathcal{A}^\kappa_{LL'JJ'}\sumint_{n'}\gamma^{(\kappa)}_{n'L'SJ'}\mathcal{R}_{nLJ,n'L'J'}^2\nonumber
\end{align}
It is composed of an angular term $\mathcal{A}^\kappa_{LL'JJ'}$ defined in (\ref{eq:AngularTerm}), a radial term $\mathcal{R}_{nLJ,n'L'J'}$ defined in (\ref{eq:RadialTerm}), and a resonance term $\gamma^{(\kappa)}_{q}$ for the transition to the virtual level $\ket{q}\equiv \ket{n'L'SJ'}$ whose original expression (\ref{eq:GammaOld})
\begin{align}
\gamma^{(\kappa)}_q(\omega)=\frac{2\omega_q}{\hbar(\omega^2-\omega_q^2)}\left(\delta_{\kappa,\text{even}}+\frac{\omega}{\omega_q}\delta_{\kappa,1}\right)\nonumber
\end{align}
we replace with (\ref{eq:GammaNew})
\begin{align}
    \gamma^{(\kappa)}_q(\omega)&=\frac{2\omega_q}{\hbar(\omega^2-\omega_q^2)}\left(\frac{\omega_q}{\omega}\right)^\kappa\nonumber
\end{align}
where $\omega_q$ is the angular transition frequency for the virtual transition. Observe how the modification to the resonance term to achieve numeric stability is quite minor.
This reduced matrix element is related to the scalar, vector, and tensor polarizabilities in Appendix \ref{appendix:svt}.

In Fig. \ref{fig:polStack} we present the insensitive scalar, vector, and tensor polarizabilities of select Cs ground and Rydberg states using the matrix elements of the ARC python package \cite{ARC}. The ground state vector polarizability changes by $\sim 1$\%, whereas the Rydberg vector and tensor polarizabilities reduce by three orders of magnitude from the predictions of ARC and \cite{Bhowmik_VectorTrap_2025} when far detuned from narrow optical resonances. The insensitive scalar, vector, and tensor polarizabilities agree well with experimental measurements, as shown in Fig. \ref{fig:summary}.

\begin{figure}[htbp]
    \centering
    \includegraphics[width=\columnwidth]{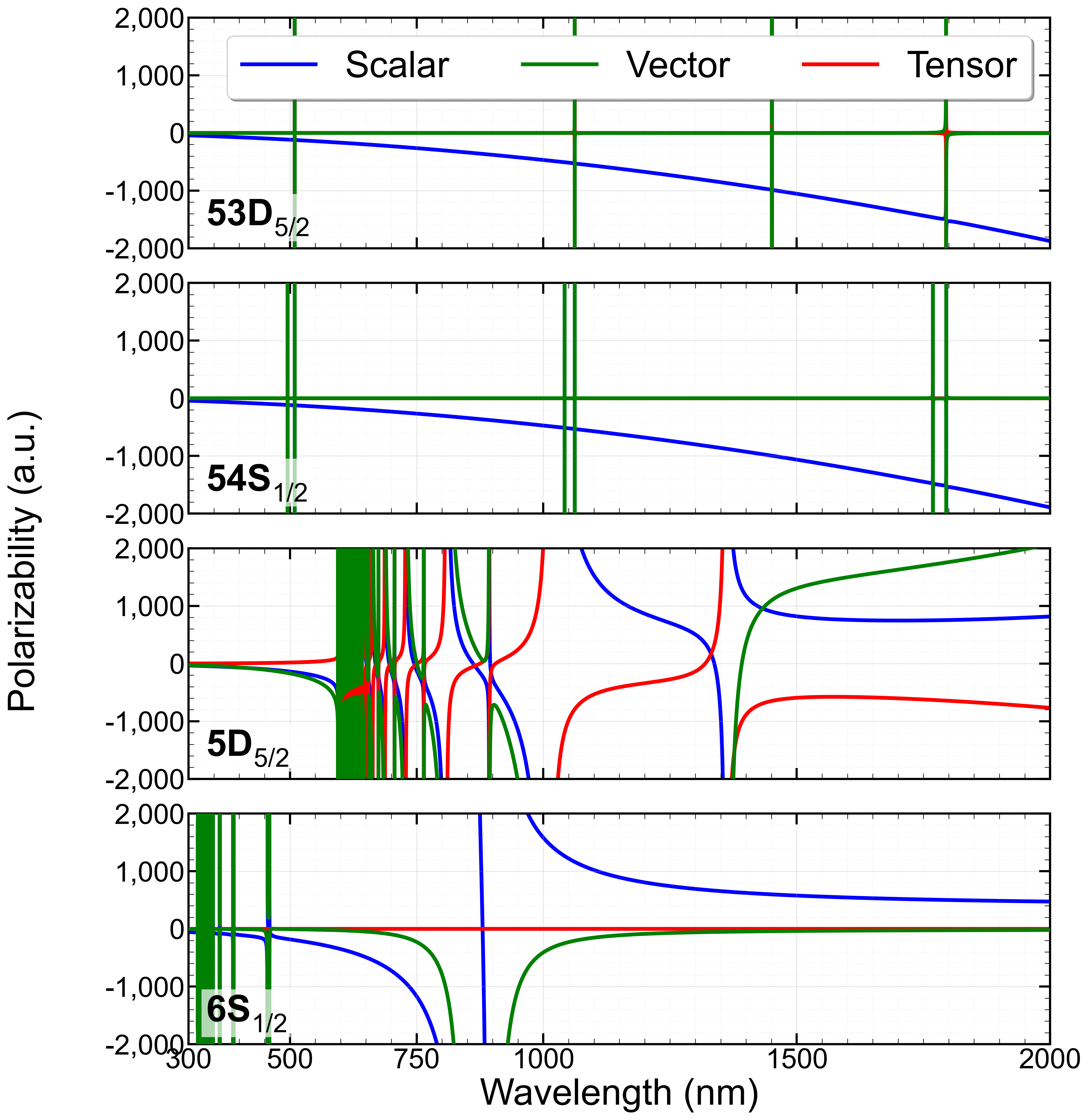}
    \caption{Predicted `insensitive' scalar, vector, and tensor polarizabilities for the $53D_{5/2}$, $54S_{1/2}$, $5D_{5/2}$, and $6S_{1/2}$ states of Cs. We compute the polarizability by subtracting the $\omega^{-1}$ and $\omega^{-2}$ dependence of the vector and tensor polarizabilities, which we prove to vanish in Sections \ref{sec:VectorPol} and \ref{sec:ScalarTensorPol}.  The polarizability is computed using a sum over bound states with radial matrix elements from the ARC python package \cite{ARC}.  We find the vector and tensor polarizabilities of Rydberg states are negligible when far detuned from optical resonances, and that their scalar polarizabilities are almost identical to the ponderomotive polarizability. The vector and tensor polarizability of non-Rydberg states experiences a several percent correction due to the modified optical frequency dependence.}
    \label{fig:polStack}
\end{figure}
\begin{figure}[htbp]
    \centering
    \includegraphics[width=\columnwidth]{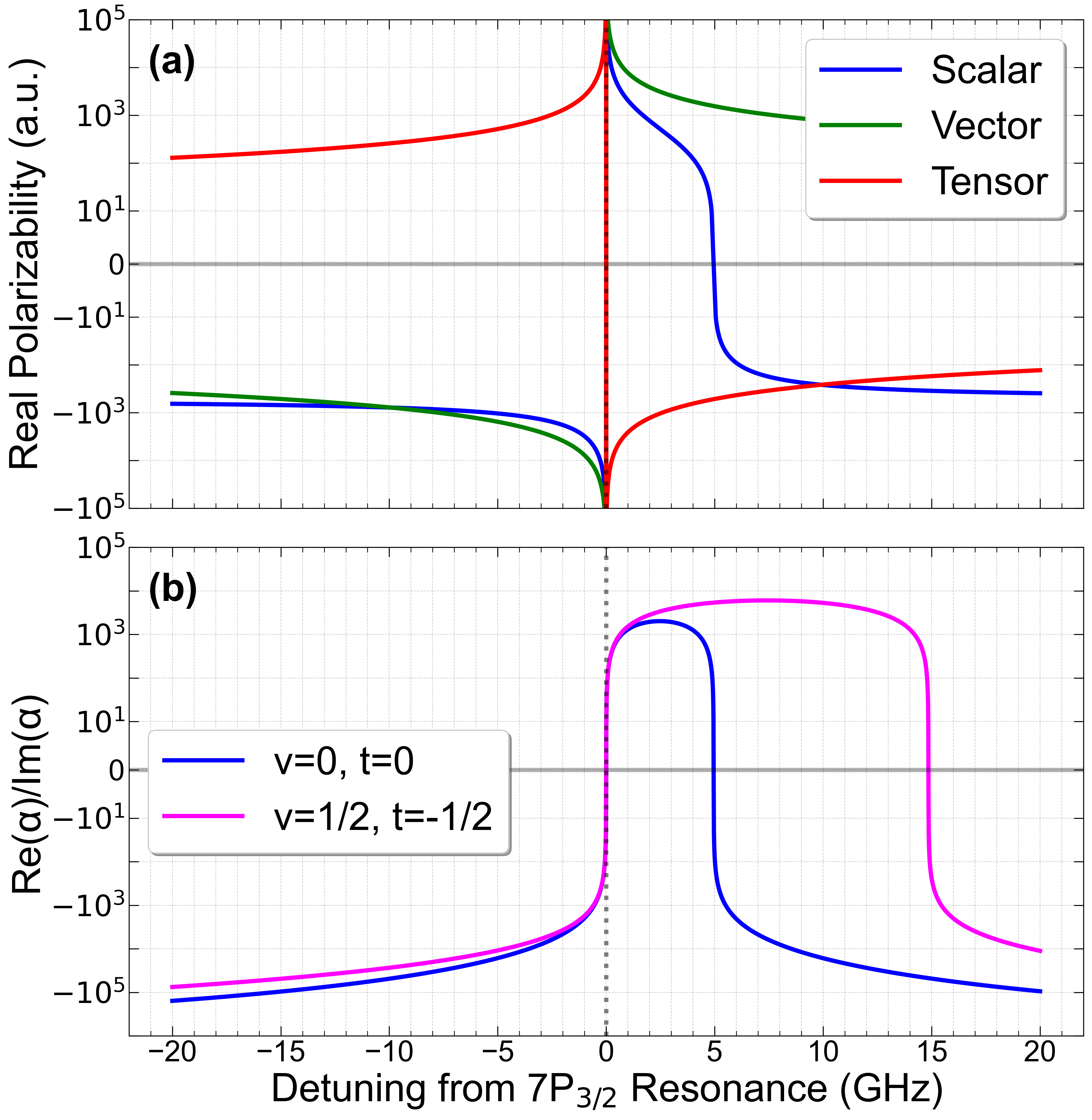}
    \caption{(a) Scalar, vector, and tensor polarizabilities of $53D_{5/2}$ near its resonance with $7P_{3/2}$ using a sum over bound states with the matrix elements from ARC \cite{ARC}. As predicted by \cite{Jansohn_Kuzmich_MagicWavelengthRydTrap_2025} the scalar polarizability is positive within several GHz blue detuning, allowing for the creation of an attractive trap.  In (b) we compute the ratio of the trap's real and imaginary polarizabilities. We  compare a trap which uses only scalar polarizability to one which uses both vector and tensor polarizabilities to maximally boost the ratio. The vector and tensor contributions, $v$ (\ref{eq:vDef}) and $t$ (\ref{eq:tDef}) and the boost $b$ (\ref{eq:bDef}) are defined in Sec. \ref{sec:NearDetuned}. The peak ratios are 2010.9 and 6021.0, respectively, at detunings of 2.466 GHz and 7.384 GHz, respectively. Their approximate ratio of $1:3$ is consistent with the analytic results of (\ref{eq:Roptb}) and (\ref{eq:deltaoptb}).}
    \label{fig:nearDetuned}
\end{figure}

Near optical resonances, the two expressions for the resonance term converge. We note that a recent proposal \cite{Jansohn_Kuzmich_MagicWavelengthRydTrap_2025} to attractively trap a Rydberg atom blue-detuned of a resonance can be enhanced by using a circularly polarized optical tweezer. As shown in Fig. \ref{fig:nearDetuned}b, for a fixed trap depth the scattering rate can be reduced threefold using a circularly polarized tweezer. Nevertheless, the reasonable trap depths the scattering lifetime is shorter than the Rydberg lifetime, suggesting this scheme is best used for short pulses intervals.

Lastly, we relax the electric-dipole assumption to consider whether a more exact atom-field interaction could be used to attractively trap an atom in the far-detuned regime. In section \ref{sec:TheoryBeyondDipole} we derive the `PZW' Hamiltonian \cite{Woolley_PZW_2020} for a Rydberg atom, which is a non-relativistically exact atom-field Hamiltonian. We use a generalized Van-Vleck high frequency expansion to prove the absence of atom-field perturbations other than the ponderomotive effect that scale as $\omega^{-1}$ or $\omega^{-2}$.

\section{Electric-dipole theory}
\label{sec:NotBeyondDipole}
\subsection{Hamiltonian}
We begin by defining the atom-field Hamiltonian and bare Hamiltonian of the atom. In optical frequencies, a Rydberg atom acts as a two particle system made from an an ionic core and a valence electron. Their mass imbalance justifies the Born-Oppenheimer approximation \cite{Born_Oppenheimer_1927}, allowing us to solve for the eigenstates of the valence electron at a relative coordinate $\vec{r}_{ec}$ with the core pinned to the position $\vec{r}_c$. In addition, the negligible spatial overlap between the valence electron and the core allows us to ignore the interaction between their spins. 

The non-relativistic atom-field interaction for these assumptions is derived in Appendix \ref{appendix:PZW}. For the electric dipole approximation, defined as an infinite wavelength, $\lambda\rightarrow \infty$ or speed of light, $c\rightarrow \infty$, the interaction becomes
\begin{align} 
V_{\text{E1}}(t) &= \left[-\vec{d}\cdot \frac{\vec{\mathcal{E}}_{\text{ac}}}{2}   e^{-i\omega t}+\text{h.c.}\right]\nonumber \\
&+ \left(- (\vec{\mu}_\text{orb}+\vec{\mu}_S) \cdot \vec{B}_{\text{dc}} -\vec{d}\cdot \vec{E}_{\text{dc}} \right.\\&\left.+ \frac{e^2 |\vec{r}_{\text{ec}} \times \vec{B}_{\text{dc}}|^2}{8m_e} \right)\nonumber \nonumber 
\end{align}
In order, the interactions are the AC electric dipole, the DC magnetic dipole, the DC electric dipole, and the DC diamagnetic effect. The dipole moments are electric $\vec{d}=-e\vec{r}_{ec}$, orbital-magnetic  $\vec{\mu}_\text{orb}=\frac{\vec{d}\times \vec{p}_{ec}}{2m_e}=-\frac{e}{2m_e}\vec{L}$, and spin-magnetic $\vec{\mu}_S=-g_S\frac{e}{2m_e}\vec{S}$. 

To first order in a relativistic correction in powers of the inverse speed of light $c^{-1}$ \cite{Leeuwen_RelativisticHydrogen_1994}, the unperturbed Hamiltonian is
\begin{align}
    H_0&= \frac{|\vec{p}|^2}{2m_e}+V(r)+H_\text{S.O.}\label{eq:HBare}
\end{align}
consisting of the valence electron's kinetic energy, a central potential $V$, and the spin-orbit interaction. The spin-orbit interaction is traditionally written as $H_\text{S.O.}=\xi(r)\vec{L}\cdot\vec{S}$ where $\xi(r)=\frac{1}{2m_e ^2c^2}\frac{1}{r}\frac{dV(r)}{dr}$\cite{BetheSalpeter1957}. For notational convenience in subsequent steps, we identify  $k_V(r)=\frac{1}{r}\frac{dV}{dr}$ as the effective spring constant of the Coulombic potential. This is possible by equating the restoring force to Hooke's law: $\vec{F}=-\vec{\nabla}V=-\left(\frac{1}{r}\frac{dV}{dr}\right)r$. The strength of the spin-orbit interaction can therefore be reformulated in terms of the ratio of the trap energy to rest energy of the electron,
\begin{align}
    H_\text{S.O.}&=\frac{\hbar\omega_V}{2}\frac{\hbar\omega_V}{m_e c^2}\frac{\vec{L}\cdot\vec{S}}{\hbar^2}
\end{align}
where $\omega_V(r)=\sqrt{k_V(r)/m_e}$ is the angular trap frequency of the Coulombic potential. 

\subsection{Perturbation Theory}
\label{sec:PolarizabilityTheory}
We consider the perturbation to a fine structure level $\mathcal{P}\equiv \ket{nLSJ}$. We define all other states to belong to the space $\mathcal{Q}$ and for $P$ and $Q$ to be the projectors of both spaces, respectively. The effective Hamiltonian for a level of energy $E_0\equiv E_{nLSJ}$ consists of a set of perturbing Hamiltonians that mix the degenerate basis of size $2J+1$. These perturbations can be expanded in powers of $V_\text{E1}$, as derived in Appendix \ref{appendix:PerturbationTheory}. We assume the detuning to states in $\mathcal{Q}$ is larger than the interaction so that the expansion rapidly converges. Indeed, the nearest fine structure states are separated by $\sim 1 h\text{GHz}$, whereas the largest atom-field interaction is the trap depth of the optical tweezer on the order of $\sim 100 h\text{MHz}$. The effective Hamiltonian to second order in $V_\text{E1}/(\hbar\Delta_0)$ becomes
\begin{align}
    H^\text{E1}_\text{eff}&=E_0 P+\delta H^{(1)}+\delta H^{(2)}_\text{dc}+\delta H^{(2)}_\text{ac}
\end{align}
where $\Delta_0$ is the angular frequency operator between the states in $\mathcal
Q$ and the states in $\mathcal{P}$.

The first-order perturbation evaluates to magnetic-dependent terms,
\begin{align}
    \delta H^{(1)}&=P\left(-(\vec{\mu}_\text{orb}+\vec{\mu}_S)\cdot\vec{B}_\text{dc}+\frac{e^2 |\vec{r}_{\text{ec}} \times \vec{B}_{\text{dc}}|^2}{8m_e}\right)P
\end{align}
and the second-order perturbation evaluates to electric-dependent terms that are separable into DC and AC parts. The DC term $\delta H^{(2)}_\text{dc}$ encodes quadratic Stark shifts, second order linear Zeeman shifts, quartic Zeeman shifts, and their cross terms. The AC term $\delta H^{(2)}_\text{ac}$ evaluates to
\begin{align}
H^{(2)}_\text{ac}&=P\left(\frac{\vec{\mathcal{E}}^*_\text{ac}}{2}\cdot\vec{d}R^{(0)}_{-1}\vec{d}\cdot\frac{\vec{\mathcal{E}}_\text{ac}}{2}+\frac{\vec{\mathcal{E}}_\text{ac}}{2}\cdot\vec{d}R^{(0)}_{+1}\vec{d}\cdot\frac{\vec{\mathcal{E}}^*_\text{ac}}{2}\right)P
\end{align}
where we define the resolvent and internal transition angular frequency operators to be
\begin{align}
    R^{(0)}_k&=-\frac{Q}{\hbar\Delta_k}\label{eq:resolvent}\\
    \hbar\Delta_k&=Q(k\hbar\omega + H^{(0)})Q-E_0\label{eq:transitionAngularFrequencyOperator}
\end{align}
The $k\hbar\omega$ term is due to the $e^{ik\omega t}$ terms of $V_\text{E1}$, which raises the energy of the intermediate state by $k\hbar\omega$ due to the emission of $k$ photons. A more rigorous explanation is provided in Appendix \ref{appendix:FloquetPhoton} using Floquet-photon theory. A correction to the resolvent that accounts for the finite lifetime of the states in $\mathcal{Q}$ is provided in Appendix \ref{appendix:FiniteLifetime}; there we prove the correction is negligible and does not affect the subsequent results for the far-detuned regime.

The two  `co-rotating' and `counter-rotating' terms of $H^{(2)}_\text{ac}$ combine with the following spherical tensor identities \cite{Zare1988}:
\begin{align}
    (\vec{E}_1\cdot \vec{I}_1)(\vec{I}_2\cdot\vec{E}_2)
    &=\sum_{\kappa=0}^2(-1)^\kappa\left(\left[\vec{I}_1\otimes\vec{I}_2\right]^\kappa\cdot\left[\vec{E}_1\otimes\vec{E}_2\right]^\kappa\right)\\
    \left[\vec{E}_1\otimes\vec{E}_2\right]^\kappa&=(-1)^\kappa\left[\vec{E}_2\otimes\vec{E}_1\right]^\kappa\\
    I^{(\kappa)}\cdot E^{(\kappa)}&=\sum_{q=-\kappa}^\kappa (-1)^q I^{(\kappa)}_q E^{(\kappa)}_{-q}
\end{align}
The AC perturbation is thus separable into a polarizability ($\alpha$) operator and a field ($\mathcal{F}$) operator:
\begin{align}
\delta H^{(2)}_\text{ac}&=\sum_{\kappa=0}^2 (-1)^\kappa\alpha^\kappa \cdot\mathcal{F}^\kappa
\label{eq:secondOrderAF}
\end{align}
The field operator is independent of the internal state of the atom. It is the decomposition of the dyadic of the vectors $\vec{\mathcal{E}}^*$ and $\vec{\mathcal{E}}$ into the following irreducible spherical tensors of rank $\kappa$, which we derive in Appendix  \ref{appendix:CartesianSpherical}
:
\begin{align}
\mathcal{F}^{(\kappa)}&=\left[\frac{\vec{\mathcal{E}}^*}{2}\otimes\frac{\vec{\mathcal{E}}}{2}\right]^{(\kappa)}\nonumber\\
&\cong\frac{1}{4}\left(-\frac{\vec{\mathcal{E}}^*\cdot\vec{\mathcal{E}}}{\sqrt{3}}\delta_{\kappa,0}+\frac{i\vec{\mathcal{E}}^*\times\vec{\mathcal{E}}}{\sqrt{2}}\delta_{\kappa,1}\right.\nonumber\\
&\left.+\left[\frac{\vec{\mathcal{E}}^*\otimes\vec{\mathcal{E}}+\vec{\mathcal{E}}\otimes\vec{\mathcal{E}}^*}{2}-\frac{\vec{\mathcal{E}}^*\cdot\vec{\mathcal{E}}}{3}\mathcal{I}\right]\delta_{\kappa,2}\right)\label{eq:FieldTensor}\end{align}
The polarizability operator is defined similarly as the rank-$\kappa$ projection of the dyadic formed by two internal vectors.
\begin{align}
\alpha^{(\kappa)}&=P\left[\vec{d}\otimes\left( R_{-1}^{(0)}+(-1)^\kappa R_{+1}^{(0)}\right)\vec{d}\right]^\kappa P\label{eq:pol_dRd}
\end{align}
The magnitude of $\alpha^{(\kappa)}$ is encoded in the reduced matrix element $\alpha^{(\kappa)}_{nLSJ}\equiv\bra{nLSJ}|\alpha^{(\kappa)}|\ket{nLSJ}$. We explain in Appendix \ref{appendix:svt} how the scalar, vector, and tensor polarizabilities are proportional to $\alpha^{(\kappa)}_{nLSJ}$ for $\kappa\in\{0,1,2\}$, respectively. 

\subsection{Decomposition of the Resolvent}
\label{sec:DecompResolvent}
The resolvent may be decomposed in the basis of $\mathcal{Q}$:
\begin{align}
    R^{(0)}_k&=\sum_{L'J'M'}\sumint_{n'}\frac{\ket{n'L'SJ'M'}\bra{n'L'SJ'M'}}{E^{(0)}-(E^{(0)}_{n'L'SJ'}+k\hbar\omega)}
\end{align}
All bound and continuum states in $\mathcal{Q}$ are included by summing over $n$ for bound states and integrating over energy for continuum states. We follow the notation of Ref. \cite{Shafer_PhotoionizationHydrogenic_1990} to define an effective principal quantum number for the continuum $n'=\frac{\hbar}{a_0\sqrt{2m_e E'}}$. We decompose the reduced matrix element of the polarizability in the basis of $\mathcal{Q}$ in terms of an angular term $\mathcal{A}$, a resonance term $\gamma$, and a radial term $\mathcal{R}$:
\begin{align}
    &\bra{nLSJ}|\alpha^{(\kappa)}|\ket{nLSJ}\nonumber\\&=e^2\sum_{L'J'}\mathcal{A}^\kappa_{LL'JJ'}\sumint_{n'}\gamma^{(\kappa)}_{n'L'SJ'}\mathcal{R}_{nLJ,n'L'J'}^2
    \label{eq:alphaRME}
\end{align}
The angular term is composed of $3j$ and $6j$ symbols.
\begin{align}
    \mathcal{A}^\kappa_{LL'JJ'}&=\sqrt{2\kappa+1}(-1)^{\kappa+J+J'}(2J+1)(2J'+1)\nonumber\\&\times(2L+1)(2L'+1)
    \begin{Bmatrix}1&1&\kappa\\J&J&J'\end{Bmatrix} \begin{Bmatrix}L&J&S\\J'&L'&1\end{Bmatrix}\nonumber\\&\times 
\begin{Bmatrix}L'&J'&S\\J&L&1\end{Bmatrix} 
\begin{pmatrix}L&1&L'\\0&0&0\end{pmatrix} 
\begin{pmatrix}L'&1&L\\0&0&0\end{pmatrix}\label{eq:AngularTerm}
\end{align}
the resonance terms approach $\pm\infty$ near their resonance, $\hbar\omega_q\equiv E_{q}-E_0$, and $0$ when far detuned.
\begin{align}
\gamma^{(\kappa)}_q(\omega)=\frac{2\omega_q}{\hbar(\omega^2-\omega_q^2)}\left(\delta_{\kappa,\text{even}}+\frac{\omega}{\omega_q}\delta_{\kappa,1}\right)\label{eq:GammaOld}
\end{align}
The radial term is a matrix element of the distance operator:

\begin{align}
\mathcal{R}_{nLJ,n'L'J'}&=\langle nLJ|r|n'L'J'\rangle\nonumber\\&=\int_0^\infty (r^2 dr )r R_{nLJ}(r)R_{n'L'J'}(r)\label{eq:RadialTerm}
\end{align}
 For continuum states, the radial wavefunction for hydrogen is provided by Ref. \cite{Shafer_PhotoionizationHydrogenic_1990}, however quantum defect theory must be used to define them for alkali atoms \cite{Gallagher_1994_RydbergBook}. We proceed to identify a numeric instability in the definition of the resonance terms (\ref{eq:GammaOld}).

First, we decompose the resolvent as an operator `Neumann' series \cite{Kato_PerturbationTheory_1995} in powers of $\Delta_0/\omega$ for $k\neq 0$: 
\begin{align}
    R^{(0)}_{k}&=\sum_{m=1}^\infty (-1)^m \frac{(\hbar\Delta_0)^{m-1}}{(k\hbar\omega)^m}\label{eq:ResolventNeumannExpansion}
\end{align}
This series does not necessarily converge for the optical transitions of the Rydberg atom, but remains valid if not truncated.

We introduce the Liouvillian super-operator for a closed system $\mathcal{L}(\hat{O})=\frac{i}{\hbar}[H_0,\hat{O}]$. The Liouvillian is equivalent to the time-evolution of an operator in the Heisenberg picture, such that the velocity and acceleration operators may be defined by $\vec{v}=\mathcal{L}(\vec{r})$ and $\vec{a}=\mathcal{L}^2(\vec{r})$. 
The Liouvillian of the position operator depends on the bare Hamiltonian $H_0$, defined in (\ref{eq:HBare}). 
The Liouvillian operators evaluate to the following expressions
\begin{align}
    \mathcal{L}^0(\vec{r})&=\vec{r}\\
    \mathcal{L}^1(\vec{r})&=\vec{v}=\frac{\vec{p}}{m_e}-\frac{k_V(r)r}{2m_e^2 c^2}\left(\hat{r}\times\vec{S}\right)\\
    \mathcal{L}^2(\vec{r})&=\vec{a}=-\frac{k_V(r)\vec{r}}{m_e}+\text{S.O. effects}
\end{align}
with a first term that is classically intuitive, followed by a relativistic correction. 

The polarizability component with $\omega^{-m}$ dependence is proportional to the operator $P\vec{d}\otimes (\hbar\Delta_0)^{m-1}\vec{d}P$. The following identity, proven in Appendix \ref{appendix:Liouvillian}, simplifies the operator in terms of the Liouvillians:
\begin{align}
    &P\vec{d}\otimes (\hbar\Delta_0)^{m}\vec{d}P=P\left(\delta_{m,\text{even}}\left\{\mathcal{L}^a(\vec{r}),\mathcal{L}^{m-a}(\vec{r})\right\}_\otimes\right.\nonumber\\&\left.+\delta_{m,\text{odd}}\left[\mathcal{L}^a(\vec{r}),\mathcal{L}^{m-a}(\vec{r})\right]_\otimes\right)P(-1)^{m-a}\frac{e^2}{2}(i\hbar)^m 
    \label{eq:LiouvillianIdentity}
\end{align}
where $a$ is any integer between 0 and $m$. We introduce the shorthand notation for the multiplication rule used within the commutator and anti-commutators, such that $[\vec{A},\vec{B}]_*=\vec{A}*\vec{B}-\vec{B}*\vec{A}$ where $*$ can represent a dot product ($\cdot$), cross product ($\times$), or outer product $(\otimes)$. 

The polarizability thus becomes a power series of $\omega^{-1}$ whose leading order terms well-approximate the effect of an AC field on an alkali Rydberg atom. Defining the anti-commutator $A^{(m)}_{a*}=\left\{\mathcal{L}^a(\vec{r}),\mathcal{L}^{m-1-a}(\vec{r})\right\}_*$ and commutator $C^{(m)}_{a*}\equiv \left[\mathcal{L}^a(\vec{r}),\mathcal{L}^{m-1-a}(\vec{r})\right]_*$, the polarizability becomes
\begin{equation}
\begin{aligned}
 &\alpha^{(\kappa)} = \sum_{m=1}^\infty(-1)^{a}(-i)^{m-1}\frac{e^2}{\hbar\omega^m}\\&\times P\left[\delta_{\kappa,1}\delta_{m,\text{odd}}A^{(m)}_{a\otimes}+\delta_{\kappa,\text{even}}\delta_{m,\text{even}}C^{(m)}_{a\otimes}\right]^{(\kappa)}P
\end{aligned}
\end{equation}
We further simplify by evaluating the rank-$\kappa$ projection of the dyadic using the results of Appendix \ref{appendix:CartesianSpherical}:
\begin{align}
  &\alpha^{(\kappa)} \cong \sum_{m=1}^\infty (-1)^{a}(-i)^{m-1}\frac{e^2}{\hbar\omega^m}\nonumber\\&\times P\left[-\frac{C^{(m)}_{a\cdot}}{\sqrt{3}}\delta_{\kappa,0}\delta_{m,\text{even}}+\frac{iA^{(m)}_{a\times}}{\sqrt{2}}\delta_{\kappa,1}\delta_{m,\text{odd}}\right.\nonumber\\
 &\left.+\left(\frac{C^{(m)}_{a\otimes}+(C^{(m)}_{a\otimes})^\text{T}}{2}-\frac{\text{Tr}(C^{(m)}_{a\otimes})}{3}\mathbb{I}\right)\delta_{\kappa,2}\delta_{m,\text{even}}\right]P\label{eq:alphaSimp}
\end{align}

Although the Liouvillian has complicated the definition of $\alpha^{(\kappa)}$ in terms of commutators, we gain a powerful operator formalism where each term in the expansion is the interaction of two vector operators. This allows us to evaluate the properties of each $m$-term analytically; specifically, we proceed to rigorously demonstrate that the $\omega^{-1}$ vector and $\omega^{-2}$ tensor terms vanish due to commutation symmetries, while facilitating the direct calculation of the non-zero $\omega^{-3}$ vector and $\omega^{-4}$ tensor terms.

\subsection{Vector Polarizability}
\label{sec:VectorPol}
Labeling the polarizability as $\alpha^{(\kappa)(m)}$ in terms of the rank $\kappa$ and $\omega^{-m}$ component, the largest term in equation \ref{eq:alphaSimp} appears naively to be $\alpha^{(1)(1)}$. The favorable scaling of the rank-1 polarizability was identified by Ref. \cite{Bhowmik_VectorTrap_2025} numerically as a means to overcome the ponderomotive effect, which scales as $\omega^{-2}$. However, it is analytically straightforward to show the term vanishes due to $\vec{r}\times\vec
r=0$:
\begin{align}
    \alpha^{(1)(1)}
    &\cong\frac{ie^2}{\sqrt{2}\hbar\omega}P\left\{\vec{r},\vec{r}\right\}_\times P\nonumber\\
    &=\frac{\sqrt{2}ie^2}{\hbar\omega}P\left(\vec{r}\times\vec{r}\right)P\nonumber\\
    &=0\nonumber
\end{align}
This fact may be proven without the Liouvillian formalism by examining the $m=1$ term of $R^{(0)}_k$ in (\ref{eq:ResolventNeumannExpansion}). For the $m=1$ component of the resolvent, the numerator of the rank-1 polarizability in (\ref{eq:pol_dRd}) is $[\vec{d}\otimes\vec{d}]^{(1)}$, which is isomorphic with $\vec{r}\times\vec{r}$ and must vanish.

The cancellation may be understood via the reduced matrix element of the polarizability operator (\ref{eq:alphaRME}). We consider its $\omega^{-1}$ dependence by replacing the resonance term (\ref{eq:GammaOld}) with $\frac{2}{\hbar\omega}$. We obtain
\begin{align}
    &\bra{nLSJ}|\alpha^{(1)(1)}|\ket{nLSJ}\nonumber\\&=\frac{2}{\hbar\omega}\langle nLJ|r^2|nLJ\rangle\sum_{L'J'}\mathcal{A}^\kappa_{LL'JJ'}
\end{align}

The sum $\sum_{L'J'}A_{LL'JJ'}$ can be shown numerically to vanish for any initial choice of $LJ$. Thus, the cancellation of the $\omega^{-1}$ term is due to destructive interference of angular momentum channels. Each angular momentum channel yields a immense vector polarizability. We speculate that a small, systematic error in the matrix elements used by Ref. \cite{Bhowmik_VectorTrap_2025} and the python package ARC \cite{ARC} is responsible for producing a large vector polarizability out of the difference of multiple immense terms that should cancel.

The leading order vector polarizability must therefore originate from a higher power of $\omega^{-1}$. We calculate $\alpha^{(1)(3)}$ for the choice $a=1$, and find it to be proportional to the cross product of the velocity operator with itself.
\begin{align}
    \alpha^{(1)(3)}&\cong-\frac{ie^2}{\sqrt{2}\hbar\omega^3}PA^{(3)(1)}_\times P\nonumber\\
    &=-\frac{\sqrt{2}ie^2}{\hbar\omega^3} P\left(\vec{v}\times\vec{v}\right)P\nonumber\\
    &\neq 0\label{eq:vecCancellation}
\end{align}
Unlike the term $\vec{r}\times\vec{r}=0$, the cross product $\vec{v}\times\vec{v}$ does not cancel due to the non-commutativity of the velocity components arising from the spin-orbit interaction. We find
\begin{align}
    \alpha^{(1)(3)}&\cong-\frac{e^2}{\sqrt{2}m_e^3c^2\omega^3}\left[(2k_V+rk_V')\vec{S}\right.\nonumber\\&\left.-\left(rk_V'+\frac{r^2k_V^2}{2m_e c^2}\right)(\hat{r}\cdot\vec{S})\hat{r}\right]
\end{align}
A numeric evaluation of this term requires an accurate alkali potential $V(r)$. Instead, having found this term to be nonzero, we provide an improved expression for a sum over states calculation of the reduced matrix element $\alpha^{(\kappa)}_{nLSJ}$. We redefine the resonance coefficient $\gamma^{(1)}_q(\omega)$ in equation \ref{eq:GammaOld} by subtracting its erroneous $\omega^{-1}$ term:
\begin{align}
\gamma^{(1)}_q(\omega)&=\frac{2\omega}{\hbar(\omega^2-\omega_q^2)}-\left(\frac{2}{\hbar\omega}\right)\nonumber\\
&=\frac{2\omega_q}{\hbar(\omega^2-\omega_q^2)}\frac{\omega_q}{\omega}\label{eq:GammaNewVector}
\end{align}
This expression is exact even when the ratio $\omega_q\ll\omega$ is not assumed, and it is less sensitive to small systematic errors in the model potential used by numeric calculations. The predicted polarizability using \ref{eq:GammaNewVector} is provided in Fig. \ref{fig:polStack}.

\subsection{Scalar and Tensor Polarizability}
\label{sec:ScalarTensorPol}
The leading order scalar and tensor polarizabilities are proportional to $\omega^{-2}$ and defined by
\begin{align}
     \alpha^{(0)(2)}&\cong\frac{ie^2}{\sqrt{3}\hbar\omega^2}PC^{(2)(0)}_\cdot P\\
      \alpha^{(2)(2)}&\cong-\frac{ie^2}{\hbar\omega^2} P\left(\frac{C^{(2)(0)}_\otimes+(C^{(2)(0)}_\otimes)^\mathrm{T}}{2}\right.\nonumber\\&\left.-\frac{\text{Tr}(C^{(2)(0)}_\otimes)}{3}\mathbb{I}\right)P
\end{align}
The commutator $C^{(2)(0)}_* =[\vec{r},\vec{v}]_*$ simplifies by noting that both $\vec{r}$ and the spin-orbit term have no momentum dependence, and therefore commute. The identities $[\vec{r},\vec{p}]_\cdot=3i\hbar$ and $[\vec{r},\vec{p}]_\otimes=i\hbar\mathbb{I}+\vec{r}\wedge\vec
p$ further simplify the commutators such that the polarizabilities become
\begin{align}
     \alpha^{(0)(2)}\cong&-\frac{\sqrt{3}e^2}{m_e\omega^2}P\nonumber\\
     =&-\sqrt{3}\alpha_P P\\
      \alpha^{(2)(2)}=&-\frac{ie^2}{\hbar m_e\omega^2}P\left(\frac{i\hbar\mathbb{I}-\vec{r}\wedge\vec
p+i\hbar\mathbb{I}+\vec{r}\wedge\vec
p}{2}\nonumber\right.\\&\left.-\frac{\text{Tr}(i\hbar\mathbb{I}+\vec{r}\wedge\vec
p)}{3}\mathbb{I}\right)P\nonumber\\
=&-\frac{ie^2}{\hbar m_e \omega^2}P\left(i\hbar\mathbb{I}-i\hbar\mathbb{I}\right)P\nonumber\\
=&\,0\label{eq:tenCancellation}
\end{align}
Observe how the $\omega^{-2}$ dependence of the rank-2 polarizability vanishes, whereas the rank-0 term is proportional to the ponderomotive polarizability $\alpha_\text{pond}=-\frac{e^2}{m_e\omega^2}$. All together, the leading order AC perturbation in powers of $\omega^{-1}$ reproduces the ponderomotive interaction:
\begin{align}
\delta H^{(2)(2)}&=\alpha^{(0)(2)}\cdot\mathcal{F}^{(0)}\nonumber\\&=-\alpha_\text{pond}P\left|\frac{\vec{\mathcal{E}}}{2}\right|^2
\end{align}
The signs show the ponderomotive effect is repulsive for higher intensities, and the $P$ shows that the ponderomotive effect affects all states in the level $\ket{nLSJ}$ equally without changing their composition. We note that $\alpha^{(0)(2)}$ has been similarly evaluated using the `TRK' sum \cite{Topcu_DynamicPolarizabilityLengthVelocityGauge_2013}.

The first nonzero rank-2 contribution has $\omega^{-4}$ dependence, and is given by the commutator $C^{(2)(1)}_\otimes=[\vec{v},\vec{a}]_\otimes$. Ignoring spin orbit effects, the leading order tensor polarizability evaluates to
\begin{align}
    \alpha^{(2)(4)}&=\frac{e^2}{m_e^2\omega^4}P\left(r\frac{dk_V(r)}{dr}\right)[\hat{r}\otimes\hat{r}]^{(2)}P+\text{S.O. terms}
\end{align}
The dependence on the derivative of the effective spring constant reveals that the non-relativistic tensor polarizability measures the anharmonicity of the potential.

Here too, we improve the expression for the resonance term $\gamma^{(2)}_q(\omega)$ by subtracting its erroneous $\omega^{-2}$ term. We provide an expression for the resonance terms valid for all rank $\kappa$:
\begin{align}
    \gamma^{(\kappa)}_q(\omega)&=\frac{2\omega_q}{\hbar(\omega^2-\omega_q^2)}\left(\frac{\omega_q}{\omega}\right)^\kappa\label{eq:GammaNew}
\end{align}
A comparison plot for the predicted tensor polarizability using this expression is provided in Figure \ref{fig:polStack}.

The new resonance term (\ref{eq:GammaNew})
is valid in all regimes, except for the DC limit $\omega\rightarrow 0$ in which it diverges. In contrast, the original resonance term (\ref{eq:GammaOld}) can be used in the DC limit to find the DC polarizability. Nevertheless, we find excellent agreement between the new and old expression for frequencies as low as $1 $GHz. We note that both expressions produce similar results near resonances due to the correction factor, $\omega_q/\omega$, being close to one.
\section{Near Detuned Vector Trapping}
\label{sec:NearDetuned}
It has recently been proposed \cite{Jansohn_Kuzmich_MagicWavelengthRydTrap_2025} to attractively trap Rydberg atoms near blue detuned of an optical transition with a lower lying excited level. Here, we quantify the quality of the trap by analyzing its dependence on polarization and calculating the associated scattering rate. 

In Appendix \ref{appendix:FiniteLifetime} we compute the numerically stable resonance terms (\ref{eq:GammaGamma0})-(\ref{eq:GammaGamma2}) by accounting for the finite lifetime $2\pi/\Gamma$ of the states in $\mathcal{Q}$.  We define being near detuned to a particular level $\ket{q'}\equiv\ket{n'L'SJ'}$ by the detuning $\delta_{q'}=\omega-|\omega_{q'}|\ll\omega$ being small but nevertheless outside of resonance: $\Gamma_{q'}\ll |\delta_{q'}|\ll\omega$. The resonance term that is near-detuned simplifies to
    \begin{align}
    \gamma^{(\kappa)}_{q'}=[\text{sgn}(\omega_{q'})]^{\kappa+1}\left[\frac{1}{\hbar\delta_{q'}}-i\frac{\Gamma_{q'}/2}{\hbar\delta_{q'}^2}\right]\label{eq:GammaNearDetuned}
\end{align}
We note that (\ref{eq:GammaNearDetuned}) is not affected by the subtraction of the `zero terms', because (\ref{eq:GammaOld}) and (\ref{eq:GammaNew}) are equivalent near resonances.

Unlike the far-detuned regime ($\delta_{q'}\approx\omega)$, the vector and tensor polarizabilities can exceed the scalar polarizability. We find the ratio of the complex vector and tensor polarizabilities using only the dominant resonance term to be 
\begin{align}
    \frac{\alpha^v}{\alpha^s}&=-\text{sgn}(\omega_{q'})\frac{3X}{2(J+1)}\label{eq:VSratio}\\
    \frac{\alpha^t}{\alpha^s}&=-\frac{3X(X-1)-8J(J+1)}{2(J+1)(2J+3)}\label{eq:TSratio}
\end{align}
where we define $X\equiv J(J+1)+2-J'(J'+1)$. The selection rules constrain $J'\in \{J-1,J,J+1\}$, and we find the greatest ratio for $J'=J-1$ such that $\left|\frac{\alpha^v}{\alpha^s}\right|=3$.

The quality of a trap is determined by its angular scattering rate $\Gamma_\text{sc}$ from a fixed trap of depth, $U_0=-\delta H^{(2)}$ (c.f. \ref{eq:svtDef}):
\begin{align}
    \hbar\Gamma_\text{sc}&=2\frac{U_0}{R}
\end{align}
where $R$ is the ratio of the real and imaginary polarizabilities: $R\equiv\text{Re}(\alpha)/\text{Im}(\alpha)$. We separate the ratio into its near-detuned and background contributions, assuming the background imaginary polarizability is negligible:
    \begin{align}
       R&\approx\frac{\text{Re}(\alpha_{q'})+\text{Re}(\alpha_\text{bg})}{\text{Im}(\alpha_{q'})}\\&=\frac{2\delta_{q'}}{\Gamma_{q'}}-\frac{|\alpha_\text{pond}|}{\text{Im}(\alpha_{q'})}
    \end{align}
where we note that the ponderomotive polarizability $\alpha_\text{pond}$ is negative. The first term is that of a two level system, and is unaffected by the trap polarization. However, the presence of a scalar background polarizability allows polarization to affect $R$ via the term $\text{Im}(\alpha_{q'})$. To quantify this, we relate the imaginary polarizability of the resonance to its scalar component by defining a polarization-dependent `boost' factor $b$:
\begin{align}
    b&=\frac{\text{Im}(\alpha_{q'})}{\text{Im}(\alpha^s_{q'})}\\&=1-\text{sgn}(\omega_{q'})v \frac{3X}{2(J+1)}-t\frac{3X(X-1)-8J(J+1)}{2(J+1)(2J+3)}\label{eq:bDef}\\
    v&=c\frac{M}{2J}\label{eq:vDef}\\
    t&=-d\frac{3M^2-J(J+1)}{2J(2J-1)}\label{eq:tDef}
\end{align}
where $\alpha^s_{q'}$ is the scalar polarizability computed using only the $q'$ resonance term. We define $v$ and $t$ \cite{Kien2013Polarizability} to be the contribution of vector and tensor effects, respectively. The terms $\text{Im}(\alpha^s_{q'})$ and $\text{Im}(\alpha_{q'})$ are strictly positive, such that $R$ is maximized by choosing polarizations ($c$ and $d$) and states $\ket{J,M}$ to maximize $\text{Im}(\alpha_{q'})$. The factor $b\in (0,3]$ is maximized for a stretched state $\ket{J,J}$ near a transition with $J'=J-1$ using $c=-\text{sgn}(\omega_{q'})$ and $d=1$, in which case $\text{Im}(\alpha_{q'})=3\text{Im}(\alpha^s_{q'})$. Using $\text{Re}(\alpha^s)=\text{Re}(\alpha^s_{q'})-|\alpha_\text{pond}|$ and $\text{Re}(\alpha^s_{q'})/\text{Im}(\alpha^s_{q'})=2\delta_{q'}/\Gamma_{q'}$ we can express the ratio entirely in terms of the scalar polarizability as
\begin{align}
    R(b)&=\frac{\text{Re}(\alpha^s)+\frac{b-1}{b}|\alpha_\text{pond}|}{\text{Im}(\alpha^s)}
\end{align}
For positive detunings the ratio is initially positive, then approaches negative infinity as the background dominates. We find the largest detuning at which a trap can be formed by solving for $R=0$. This occurs when
\begin{align}
    \text{Re}(\alpha^s)\big|_{R=0}=-\frac{b-1}{b}|\alpha_\text{pond}|
\end{align}
Thus, we conclude that a vector enhanced trap with $b=3$ can extend the trapping region to regions where the scalar polarizability is negative, up to $\text{Re}(\alpha^s)=-\frac{2}{3}|\alpha_\text{pond}|$. We also solve for the detuning at which $R$ is maximized with respect to $\delta_{q'}$ to find that the optimal detuning $\delta_0(b)=\delta_{q'}\big|_{dR(b)/d\delta=0}$ is pushed out by $b$
\begin{align}
    \delta_0(b)=b \delta_0(1)\label{eq:deltaoptb}
\end{align}
and that at this optimal detuning, the scalar polarizability is defined by
\begin{align}
    \text{Re}(\alpha^s(\delta_0(b)))&=\frac{2-b}{b}|\alpha_\text{pond}|
\end{align}
Evaluating $R$ at this detuning results in a clear expression
\begin{align}
    \max(R(b))=b\max(R(1))=\frac{1}{2}\frac{2\delta_0(b)}{\Gamma_{q'}}\label{eq:Roptb}
\end{align}
Thus, the quality of the optimal trap is improved by $b$, although the quality is at best half that of the two level system without any background. The sole advantage of a larger $b$ is to increase the optimal detuning since the real and imaginary polarizabilities scale as $\delta^{-1}$ and $\delta^{-2}$, respectively. 

In Fig. \ref{fig:nearDetuned}b we demonstrate agreement between these results on the perturbation of $53D_{5/2}$ by the level $7P_{3/2}$. We note that because $\text{max}(R(b))$ is linked to that of a two-level system, the performance of near detuned trapping can be enhanced by reducing $\Gamma_{q'}$ and increasing $\text{Re}(\alpha^s_{q'})$. An intermediate level $nP_{3/2}$ with a larger principal quantum number $n$ would do so, at the cost of increasing the trap wavelength.

\section{Beyond Dipole}
\label{sec:TheoryBeyondDipole}
\subsection{Exact Atom-Field Hamiltonian}
\label{subsec:ExactAtomFieldHamiltonian}
As discussed in section \ref{sec:Definition}, the value of $k r_{rms}$ exceeds 1 for the measured Rydberg states. The electric dipole approximation assumes a uniform field across the atom. However, the variation of the field can drive higher multipole moments. We investigate whether the dominant perturbation from the atom-field interaction remains the ponderomotive effect in this regime.

In Appendix \ref{appendix:PZW} we adapt the formalism of Power, Zienau, and Woolley \cite{Woolley_PZW_2020} to an alkali Rydberg atom and obtain the following exact non-relativistic atom-field interaction:
\begin{align}
    V_\text{PZW}=&-\vec{\mu}_\text{orb}\cdot\langle\vec{B}_\text{ext}\rangle-\vec{\mu}_S\cdot \vec{B}_\text{ext}(\vec{r}_e)\\&+\frac{|\vec{d}\times\langle\vec{B}_\text{ext}\rangle|^2}{8m_e}-\vec{d}\cdot\langle\vec{E}_\text{ext}\rangle
\end{align}
Similar to $V_\text{E1}$, this interaction includes the linear Zeeman effect, the diamagnetic effect, and electric-dipole interaction. However, the external fields experienced by the atom are not local; they are weighted averages along the straight line between the core and the valence electron:
\begin{align}
    \langle \vec{E}_\text{ext}\rangle&=\int_0^1 d\lambda\, \vec{E}_\text{ext}\left(\vec{r}_c+\lambda\vec{r}_{ec}\right)\\
        \langle \vec{B}_\text{ext}\rangle&=2\int_0^1 d\lambda\, \lambda\,\vec{B}_\text{ext}\left(\vec{r}_c+\lambda\vec{r}_{ec}\right)
\end{align}
Their non-locality is insignificant when the atom is smaller than the variation of the fields, in which case $V_\text{PZW}$ simplifies to $V_\text{E1}$. These integrals are a closed form of the multipole expansion that can be obtained by a Taylor series about $\vec{r}_c$. 

We note that the minimal coupling \cite{Cohen-Tannoudji_PhotonsAndAtoms_1989:} and Kramers-Henneberger \cite{Henneberger_KHHamiltonianIntenseLight_1968} Hamiltonians can equivalently describe the atom-field interaction in terms of gauge-dependent potentials. An advantage of the `PZW' Hamiltonian is that its terms are observables and the frame of reference is inertial.

We assume the external fields are composed of static, uniform fields $\vec{E}_\text{dc}$ and $\vec{B}_\text{dc}$, as well as a spatially varying monochromatic optical tweezer of the following form:
\begin{align} 
\vec{E}_{\text{ac}}(\vec{r},t)&=\Re\left(\vec{\mathcal{E}}_{\text{ac}}(\vec{r})e^{-i\omega t}\right) \\ 
\vec{B}_{\text{ac}}(\vec{r},t)&=\Re\left(\vec{\mathcal{B}}_{\text{ac}}(\vec{r})e^{-i\omega t}\right) 
\end{align}
In addition, we use Faraday's law to rewrite the AC magnetic field in terms of a curl, $\vec{\mathcal{C}}_\text{ac}$, that has the same dimension as $\vec{\mathcal{E}}_\text{ac}$:
\begin{align}
\vec{\mathcal{B}}_{\text{ac}} &= -\frac{i}{c}\vec{\mathcal{C}}_\text{ac}\\
\vec{\mathcal{C}}_\text{ac}&\equiv\frac{\vec{\nabla}}{k} \times \vec{\mathcal{E}}_{\text{ac}}
\end{align}
The exact atom-field interaction is thus
\begin{align} 
V_\text{PZW}&=\left[-\frac{(\vec{d}\times\langle\vec{\mathcal{C}}_{\text{ac}}\rangle)^2}{32m_e c^2}e^{-2i\omega t} +\text{h.c.}\right]\nonumber \\ 
&+\left[\left(\frac{i\vec{\mu}_{\text{orb}}}{c}\cdot\frac{\langle \vec{\mathcal{C}}_{\text{ac}}\rangle}{2}+\frac{i\vec{\mu}_S}{c}\cdot\frac{\vec{\mathcal{C}}_{\text{ac}}(\vec{r}_e)}{2}-\vec{d}\cdot\frac{\langle\vec{\mathcal{E}}_{\text{ac}}\rangle}{2}\right.\right.\nonumber\\
&\left.\left.-i\frac{(\vec{d}\times\vec{B}_{\text{dc}})\cdot(\vec{d}\times\langle\vec{\mathcal{C}}_{\text{ac}}\rangle)}{8m_e c}\right)e^{-i\omega t} +\text{h.c.}\right]\nonumber \\ 
&+\left[-(\vec{\mu}_{\text{orb}}+\vec{\mu}_S)\cdot \vec{B}_{\text{dc}}-\vec{d}\cdot\vec{E}_{\text{dc}}\right.\nonumber\\& \left.+ \frac{|\vec{d}\times\vec{B}_{\text{dc}}|^2}{8m_e} + \frac{|\vec{d}\times\langle\vec{\mathcal{C}}_{\text{ac}}\rangle|^2}{16m_e c^2} \right]\label{eq:PZWAF}
\end{align}
\subsection{High Frequency Expansion}

We use the Van Vleck high frequency perturbation theory to rigorously decompose the effective Hamiltonian in powers of $1/\omega$ to determine the significance of beyond-dipole terms.

The effective Van Vleck Hamiltonian is derived in Appendix \ref{appendix:VV} to be
\begin{align}
    &H_\text{VV}=H^{(0)}_0+\left(H^{(1)}_0+\sum_{n\neq 0}\frac{[H^{(0)}_n,H^{(0)}_{-n}]}{2n\hbar\omega}\right)\nonumber\\&+\left(-\sum_{n\neq 0}\frac{1}{2n^2\hbar^2\omega^2}[H^{(0)}_n,[H^{(0)}_{-n},H^{(0)}_0]]\nonumber\right.\\&\left.+\sum_{\substack{n,m\neq 0\\n+m\neq 0}}\frac{1}{3nm\hbar^2\omega^2}[H^{(0)}_n,[H^{(0)}_m,H^{(0)}_{-n-m}]]\right.\nonumber\\
    &+\left. \sum_{n\neq 0}\frac{[H^{(0)}_n,H^{(1)}_{-n}]}{n\hbar\omega}+H^{(2)}_0\nonumber\right)+\mathcal{O}(\omega^{-3})
\end{align}
where $H^{(m)}_k$ is the Fourier component of the Hamiltonian proportional to $\omega^{-m}$ with time dependence  $e^{ik\omega t}$. 

We use the full Hamiltonian $H=H_0+V_\text{PZW}$ to solve for $H_\text{VV}$:

\begin{align} 
H&=\left[-\frac{(\vec{d}\times k\langle\vec{\mathcal{C}}_{\text{ac}}\rangle)^2}{32m_e \omega^2}e^{-2i\omega t} +\text{h.c.}\right]\nonumber \\ 
&+\left[\left(\frac{ik\vec{\mu}_{\text{orb}}}{\omega}\cdot\frac{\langle \vec{\mathcal{C}}_{\text{ac}}\rangle}{2}+\frac{ik\vec{\mu}_S}{\omega}\cdot\frac{\vec{\mathcal{C}}_{\text{ac}}(\vec{r}_e)}{2}-\vec{d}\cdot\frac{\langle\vec{\mathcal{E}}_{\text{ac}}\rangle}{2}\right.\right.\nonumber\\
&\left.\left.-i\frac{(\vec{d}\times\vec{B}_{\text{dc}})\cdot(\vec{d}\times k\langle\vec{\mathcal{C}}_{\text{ac}}\rangle)}{8m_e \omega}\right)e^{-i\omega t} +\text{h.c.}\right]\nonumber \\ 
&+\left[-(\vec{\mu}_{\text{orb}}+\vec{\mu}_S)\cdot \vec{B}_{\text{dc}}-\vec{d}\cdot\vec{E}_{\text{dc}}+ \frac{|\vec{d}\times\vec{B}_{\text{dc}}|^2}{8m_e}\right.\nonumber\\& \left. + \frac{|\vec{d}\times k\langle\vec{\mathcal{C}}_{\text{ac}}\rangle|^2}{16m_e \omega^2}+ \frac{|\vec{p}|^2}{2m_e}+V(r)+\frac{\omega^2_V k^2}{2m_e \omega^2}\vec{L}\cdot\vec{S} \right]
\end{align}
To ensure the perturbation theory correctly sorts terms by magnitude, we make all dependence on $\omega$ explicit by substituting $c=\omega/k$. This substitution ensures that terms which physically cancel are grouped cleanly within the same order of the Van Vleck expansion. 

We find the $H_\text{VV}^{(0)}$ to be independent of the AC fields, and the $\omega^{-1}$ term to vanish due to neither $\vec{d}$ nor $\langle\vec{\mathcal{E}}_\text{ac}\rangle$ having momentum dependence:
\begin{align}
    H_\text{VV}^{(1)}&=\frac{\left[\vec{d}\cdot\langle\vec{\mathcal{E}}_\text{ac}\rangle,\vec{d}\cdot\langle\vec{\mathcal{E}}_\text{ac}^*\rangle\right]+\text{h.c.}}{8\hbar\omega}=0
\end{align}
 A lengthy derivation of $H^{(2)}_\text{VV}$ results in
\begin{align}
    H^{(2)}_\text{VV}= \frac{\omega^2_V}{2m_e c^2}\vec{L}\cdot\vec{S}+\frac{e^2|\vec{\mathcal{E}}_{\text{ac}}(\vec{r}_e)|^2}{4m_e\omega^2}
\end{align}
The result consists of a spin orbit and ponderomotive term. The spin orbit term appears due to $c=\omega/k$ and is irrelevant, not being an atom-field interaction. The ponderomotive term replicates the known beyond-dipole form that has been derived using the minimal coupling Hamiltonian \cite{Topcu_DynamicPolarizabilityLengthVelocityGauge_2013}. The electron averages the intensity over its wavefunction.  

Although the Van Vleck high-frequency expansion may diverge near optical resonances, it confirms the absence of $\omega^{-1}$ or $\omega^{-2}$ effects other than the ponderomotive effect. Furthermore, optical transition matrix elements depend on the spatial overlap with a compact internal state of radius $r_\text{comp}$. Because the integrand is only non-negligible where $k r_\text{comp}\ll 1$, these optical transitions inherently satisfy the electric-dipole approximation and do not introduce higher-order beyond-dipole corrections.
\section{Conclusion}
\label{sec:Conclusion}
Through measurement and theory, we conclude that the effect of far-detuned monochromatic optical light on an alkali Rydberg atom is ponderomotive in the far-detuned regime. Although previously established \cite{Topcu_DynamicPolarizabilityLengthVelocityGauge_2013,Younge_Raithel_PonderomotiveLattice_2010}, we refine the argument by using a gauge-invariant inertial Hamiltonian and allow for interactions between AC and DC fields. In effect, we find the valence electron acts like a free particle regardless of any external fields. Thus, far-detuned monochromatic light cannot be used to form an attractive optical trap, yet conversely, it can repel all alkali Rydberg states almost equally. For an optical tweezer platform, the direct consequence is that for far-detuned light, more optical power must be expended per trap to surround the atom in light \cite{Li_Saffman_CrossedVortexBoB_2012}, and aberrations must be more rigorously controlled than for a Gaussian beam.

In addition, we develop an operator formalism to find the scalar, vector, and tensor polarizabilities of a Rydberg atom scale with the optical angular frequency $\omega$ as  $\omega^{-2}$, $\omega^{-3}$, and $\omega^{-4}$, respectively. We attribute the results of Ref. \cite{Bhowmik_VectorTrap_2025}, which predict an immense vector polarizability scaling with $\omega^{-1}$, as due to small errors in their matrix elements. We find excellent experimental and theoretical agreement for the vector polarizabilities of the $L=\{0,1,2\}$ states near $n=54$ using the refined expressions. These new expressions gently alter the predicted vector and tensor polarizabilities for lower lying states, suggesting a small systematic error has existed in all theoretically predicted vector and tensor polarizabilities. We believe the expressions found to be analytically zero can be used by theorists to benchmark their model potentials.

Lastly, we quantify the scattering rate  for near attractive trapping \cite{Jansohn_Kuzmich_MagicWavelengthRydTrap_2025}. Precisely because the background polarizability is scalar with no vector or tensor contribution,  one can control polarization and the internal state to construct an attractive trap with less scattering from the admixed level. However, for reasonable trap depths the scattering lifetime can be shorter than the bare Rydberg lifetime. It is therefore best used for short durations relative to the Rydberg lifetime, such as for `magic' excitations between the ground and Rydberg states.

\begin{acknowledgements}
  We thank Yu Wang, Ryan Cimmino, and Kenneth Wang for sharing their project's Cs Rydberg lasers. We also thank Doerte Blume, Anal Bhowmik, and Antoine Browaeys and his group for discussions. This work is supported by the U.S. Air Force Office of Scientific Research (FA9550-23-1-0538) and its Multidisciplinary University Research Initiative (FA9550-20-1-0323), as well as the Gordon and Betty Moore Foundation (DOI 10.37807/ GBMF11558).   

\end{acknowledgements}

\bibliography{Biblography/thisWorkSpecific,Biblography/master_ref_viSWAP}

\appendix 
\section{Perturbation Theory}
\label{appendix:PerturbationTheory}

A Hilbert space $\mathcal{H}$ can be separated into two parts: $\mathcal{H}=\mathcal{P}\oplus\mathcal{Q}$ \cite{Lowdin_1962_PerturbationTheory}. The time independent Schrodinger equation becomes a block matrix:
\begin{align}
    \begin{pmatrix}
        E\mathbb{I}_\mathcal{P}-H_\mathcal{PP}&-H_\mathcal{PQ}\\-H_\mathcal{QP}&E\mathbb{I}_\mathcal{Q}-H_\mathcal{QQ}
    \end{pmatrix}\begin{pmatrix}
        \psi_\mathcal{P}\\\psi_\mathcal{Q}
    \end{pmatrix}=0
\end{align}
We assume the Hamiltonian contains a bare term $H_0$ that does not couple the two spaces, and a weak perturbation $V$ that does. We assume the states in $\mathcal{P}$ are degenerate with energy $E_0$. The terms in the block matrix become
\begin{align}
    H&=H_0+V\\
    H_\mathcal{PP}&=E_0 \mathbb{I}_\mathcal{P}+V_\mathcal{PP}\\
    H_\mathcal{PQ}&=V_\mathcal{PQ}\\
    H_\mathcal{QQ}&=H_{0,\mathcal{QQ}}+V_\mathcal{QQ}
\end{align}
The Schrödinger equation for states in $\mathcal{P}$ is given by
\begin{align}
    H_\text{eff}(E)\ket{\psi_\mathcal{P}}=E\ket{\psi_\mathcal{P}}
\end{align}
where we define an effective Hamiltonian and a resolvent $R$ \cite{Cohen-Tannoudji_PhotonsAndAtoms_1989:} as
\begin{align}
    H_\text{eff}(E)&=E_0\mathbb{I}_\mathcal{P}+V_\mathcal{PP}+V_\mathcal{PQ}R_\mathcal{Q}(E)V_\mathcal{QP}\\
    R_\mathcal{Q}(E)&=(E\mathbb{I}_\mathcal{Q}-H_{0,\mathcal{QQ}}-V_\mathcal{QQ})^{-1}
\end{align}
The resolvent of the unperturbed space is
\begin{align}
    R^{(0)}&=(E_0\mathbb{I}_\mathcal{Q}-H_{0,\mathcal{QQ}})^{-1}
\end{align}
The recursive operator identity
\begin{align}
    (A-B)^{-1}&=A^{-1}+A^{-1}B(A-B)^{-1}
\end{align}
can be used recursively in terms of $\delta E=E-E_0$ so that the effective Hamiltonian becomes
\begin{align}
    H_\text{eff}&=E_0\mathbb{I}_\mathcal{P}+V_\mathcal{PP}\nonumber\\&
    +V_\mathcal{PQ}\left(R^{(0)}+R^{(0)}\left[V_\mathcal{QQ}-\delta E\mathbb{I}_\mathcal{Q}\right]R^{(0)}\right)V_\mathcal{QP}\nonumber\\
    &+\mathcal{O}\left([V_\mathcal{QQ}-\delta E\mathbb{I}_\mathcal{Q}]^2\right)
\end{align}
We truncate the series to second order in the perturbation $V$. Because $\delta E$ is dependent to first order on $V_\mathcal{PP}$, the second order expansion of $H_\text{eff}$ is thus
\begin{align}
    H_\text{eff}&=E_0\mathbb{I}_\mathcal{P}+V_\mathcal{PP}+V_\mathcal{PQ}R^{(0)}V_\mathcal{QP}+\mathcal{O}(V^3)
\end{align}

\section{Definition of $\alpha^s$, $\alpha^v$, $\alpha^t$}
\label{appendix:svt}

Any two spherical tensors of the same rank, $A^{(\kappa)}$ and $B^{(\kappa)}$, can be interchanged within an angular momentum space with projector $P$ by the projection theorem \cite{Zare1988}:
\begin{align}
    P\frac{A^{(\kappa)}}{A^{(\kappa)}_{nLSJ}}P&=P\frac{B^{(\kappa)}}{B^{(\kappa)}_{nLSJ}}P
\end{align}
where we define the symmetric reduced matrix elements for the fine structure level $\ket{n(LS)J}$ by the notation $A^{(\kappa)}_{nLSJ}\equiv \bra{nJ}|A^{(\kappa)}|\ket{nJ}$.
We redefine \ref{eq:secondOrderAF} in terms of a conveniently chosen spherical tensor $\mathcal{O}^{\kappa}$ such that
\begin{align}
    P\alpha^{(\kappa)}P&=P\frac{\alpha^{(\kappa)}_{nLSJ}}{\mathcal{O}^{(\kappa)}_{nLSJ}}\mathcal{O^{(\kappa)}}P
\end{align}
The chosen spherical tensors are $\mathbb{I}^{(0)}$ for rank 0,  $J^{(1)}$ for rank 1, and $[J^{(1)}\otimes J^{(1)}]^{(2)}$ for rank 2, whose symmetric reduced matrix elements are
\begin{align}
    \langle \eta' J'||\mathbb{I}^{(0)}||\eta J\rangle &=\delta_{\eta'\eta}\delta_{J'J}\sqrt{2J+1}
\end{align}
\begin{align}
    \langle \eta' J'||J^{(1)}||\eta J\rangle &=\delta_{\eta'\eta}\delta_{J'J}\sqrt{2J+1}\sqrt{J(J+1)}
\end{align}
\begin{align}
    \langle \eta' J'||[J^{(1)}\otimes J^{(1)}]^{(2)}||\eta J\rangle &=\delta_{\eta'\eta}\delta_{J'J}\frac{1}{2\sqrt{6}}\sqrt{\frac{(2J+3)!}{(2J-2)!}}
\end{align}
The scalar, vector, and tensor polarizabilities are proportional to the ratio of these reduced matrix elements:
\begin{align}
     \alpha^s_{nLSJ}&=\frac{1}{\sqrt{3}}\frac{\alpha^{(0)}_{nLSJ}}{\langle J||\mathbb{I}^{(0)}||J\rangle}\\
    \alpha^v_{nLSJ}&=-\frac{2J}{\sqrt{2}}\frac{\alpha^{(1)}_{nLSJ}}{\bra{J}|J^{(1)}|\ket{J}}\\
     \alpha^t_{nLSJ}&=-\frac{1}{6}2J(2J-1)\frac{\alpha^{(2)}_{nLSJ}}{\langle J||[J^{(1)}\otimes J^{(1)}]^{(2)}||J\rangle}\\
\end{align}
The additional factors of $\{\frac{1}{\sqrt{3}},-\frac{2J}{\sqrt{2}},-\frac{1}{6}2J(2J-1)\}$ are included by convention. The second order perturbation becomes
\begin{align}
    \delta H^{(2)}_\text{ac}=&\sqrt{3}\alpha^s_{nLSJ}\mathbb{I}^{(0)}\cdot\mathcal{F}^{(0)}+\frac{\sqrt{2}}{2J}\alpha^v_{nLSJ}J^{(1)}\cdot\mathcal{F}^{(1)}\\&-\frac{6}{2J(2J-1)}\alpha^t_{nLSJ}[J^{(1)}\otimes J^{(1)}]^{(2)}\cdot\mathcal{F}^{(2)}
\end{align}
% We may rewrite the field tensor using the spherical basis defined in Appendix \ref{appendix:CartesianSpherical}:
% \begin{align}
% \mathcal{F}^{(0)}&=-\frac{\mathcal{E}^2}{4}\frac{1}{\sqrt{3}}\ket{0,0}_{\hat{z}}\\
% \mathcal{F}^{(1)}&=-\frac{\mathcal{E}^2}{4}\frac{c}{\sqrt{2}}\ket{1,0}_{\hat{z}}\\
% \mathcal{F}^{(2)}&=-\frac{\mathcal{E}^2}{4}\left[\frac{l}{2}\left(\ket{2,-2}_{\hat{z}}+\ket{2,2}_{\hat{z}}\right)+\frac{1}{\sqrt{6}}\ket{2,0}_{\hat{z}}\right]
% \end{align}
which evaluates to equation \ref{eq:svtDef} for the frame where $\hat{z}$ is the direction of light propagation. 
% However, if the internal angular momentum $\vec{J}$ is quantized by an external magnetic field along $\hat{Z}$, we must rotate the frame of $\mathcal{F}^{(\kappa)}$ onto magnetic field frame using Wigner-D matrix elements. 
\section{Non-Relativistic Atom-Field Interaction}
\label{appendix:PZW}

The internal charge density and current density of the Rydberg atom are 
\begin{align}
    \rho_\text{int}(\vec{r},t)&=-e\delta(\vec{r}-\vec{r}_e(t))+e\delta(\vec{r}-\vec{r}_c)\\
    \vec{j}_\text{int}(\vec{r},t)&=-e\dot{\vec{r}}_e(t)\delta(\vec{r}-\vec{r}_e(t))+(\vec{\nabla}\delta(\vec{r}-\vec{r}_e(t)))\times\vec{\mu}_S
\end{align}
where we use the magnetic dipole moment $\vec{\mu}_S\equiv g_e \frac{-e}{2m_e}\vec{S}$. These equations satisfy the continuity equation $ \vec{\nabla}\cdot\vec{j}_\text{int}+\dot\rho_\text{int}=0$. 

Let $\vec{E}$ and $\vec{B}$ denote the superposition of atom-generated and externally-generated electromagnetic fields defined in terms of gauge-dependent potentials $\vec{A}$ and $\phi$:
\begin{align}
    \vec{E} &= -\dot{\vec{A}} - \nabla \phi\\\vec{B} &= \nabla \times \vec{A}
\end{align}
The minimal-coupling Lagrangian for the atom in an electromagnetic field is given by \cite{Cohen-Tannoudji_PhotonsAndAtoms_1989:}
\begin{align}
    L_\text{MC}=&\frac{1}{2}m_e|\dot{\vec{r}}_e|^2+\frac{\epsilon_0}{2}\int d^3\vec{r}(|\vec{E}|^2-c^2|\vec{B}|^2)\\&+\int d^3\vec{r}(\vec{j}_\text{int}\cdot\vec{A}-\rho_\text{int}\phi)
\end{align}
The two integrands are chosen as the simplest relativistic invariants in the theory of electromagnetism. The two time-derivative terms, $\dot{\vec{r}}_e$ and $\dot{\vec{A}}$, define canonical particle and field momenta 
\begin{align}
    \vec{p}(\vec{r}_e)&=\frac{\partial L_\text{MC}}{\partial\dot{\vec{r}}_e}=m\dot{\vec{r}}_e-e\vec{A}(\vec{r}_e)\\
    \vec{\Pi}(\vec{r})&=\frac{\delta L_\text{MC}}{\delta\dot{\vec{A}}(\vec{r})}=-\epsilon_0\vec{E}(\vec{r})
\end{align}
in terms of which the Lagrangian may be converted into the minimal coupling Hamiltonian:
\begin{align}
    H_\text{MC}=&\vec{p}\cdot\dot{\vec{r}}_e+\int d^3\vec{r}\,\vec{\Pi}\cdot\dot{\vec{A}}-L_\text{MC}\\=&\frac{(\vec{p}+e\vec{A})^2}{2m_e}+\frac{\epsilon_0}{2}\int d^3\vec{r}\left(|\vec{E}|^2+|c\vec{B}|^2\right)\\&+\int d^3\vec{r}\,\phi\left(\rho-\epsilon_0\vec{\nabla}\cdot\vec{E}\right)-\vec{\mu}_S\cdot \vec{B}(\vec{r}_e)
\end{align}
This description is cumbersome because the fields produced externally and by the atom are mixed. Instead, we use the Power-Zienau-Woolley (PZW) Hamiltonian \cite{Woolley1971}, which partitions the fields based on their sources. 

We first define atom-generated internal fields $\vec{E}_\text{int}$ and $\vec{B}_\text{int}$ that satisfy Maxwell's equations for the atom's charge density $\rho_\text{int}$ and current density $\vec{j}_\text{int}$:
\begin{align}
    \vec{\nabla}\cdot\vec{E}_\text{int} &= \frac{\rho_\text{int}}{\epsilon_0} \\
    \vec{\nabla}\times \vec{B}_\text{int} &= \mu_0\vec{j}_\text{int}+\frac{1}{c^2}\dot{\vec{E}}_\text{int}
    \label{eq:InternalContinuity}
\end{align}
These fields are not uniquely defined because the curl of $\vec{E}_\text{int}$ is a free parameter. The internal fields may be defined by path integrals that start at the core and end at the valence electron. The freedom of the path is analogous to the gauge freedom of the electromagnetic potentials.

We define the fields applied by the experimentalist as the remainder of the total field: $\vec{E}_\text{ext} \equiv \vec{E}-\vec{E}_\text{int}$ for the electric field, and similarly for the magnetic field. These external fields satisfy the same Maxwell's equations as eqn \ref{eq:InternalContinuity} but for free charges ($\rho_\text{free}=\rho-\rho_\text{int}$) and currents ($\vec{j}_\text{free}=\vec{j}-\vec{j}_\text{int}$). 

A Lagrangian by be shifted by a total time derivative without altering the equations of motion. Observe how we can add a total time derivative to $L_\text{MC}$ to produce a new `PZW' Lagrangian \cite{Woolley_PZW_2020} that depends on the atom fields $\vec{E}_\text{int}$ and $\vec{B}_\text{int}$:
\begin{align}
    L_\text{PZW}&=L_\text{MC}+\epsilon_0\frac{d}{dt}\int d^3\vec{r}\,\vec{E}_\text{int}\cdot\vec{A}\\
    &=\frac{1}{2}m_e|\dot{\vec{r}}_e|^2+\vec{\mu}_S\cdot \vec{B}(\vec{r}_e)\\
    &+\frac{\epsilon_0}{2}\int d^3\vec{r}\left(|\vec{E}_\text{ext}|^2-|\vec{E}_\text{int}|^2-|c\vec{B}_\text{ext}|^2+|c\vec{B}_\text{int}|^2\right)
\end{align}
The canonical momenta of the PZW Lagrangian are:
\begin{align}
    \vec{p}(\vec{r}_e)&=\frac{\partial L_\text{PZW}}{\partial\dot{\vec{r}}_e}=m\dot{\vec{r}}_e+\vec{\Xi}(\vec{r}_e)\\
    \vec{\Pi}(\vec{r})&=\frac{\delta L_\text{PZW}}{\delta\dot{\vec{A}}(\vec{r})}=-\epsilon_0\vec{E}_\text{ext}(\vec{r})
\end{align}
where $\vec{\Xi}(\vec{r}_e)=\epsilon_0 c^2\vec{\nabla}_{\dot{\vec{r}}}\left(\int d^3\vec{r}\,\vec{B}_\text{int}(\vec{r})\cdot \vec{B}(\vec{r})\right)$ is the momentum of the atom's magnetic field, called the Röntgen momentum \cite{Wilkens_1994_Rontgen}.

The PZW Hamiltonian becomes
\begin{align}
    H_\text{PZW}&=\vec{p}\cdot\dot{\vec{r}}_e+\int d^3\vec{r}\,\vec{\Pi}\cdot\dot{\vec{A}}-L_\text{PZW}\\&=\frac{(\vec{p}-\vec{\Xi})^2}{2m_e}+\epsilon_0\int d^3\vec{r}\,\vec{E}_\text{int}\cdot\vec{E}_\text{ext}\\
    &+\frac{\epsilon_0}{2}\int d^3\vec{r}\,\left(|\vec{E}_\text{int}|^2+|c\vec{B}_\text{int}|^2\right)-\vec{\mu}_S\cdot \vec{B}(\vec{r}_e)+H_\text{ext}
\end{align}
where $H_\text{ext}=\frac{\epsilon_0}{2}\int d^3\vec{r}\left(|\vec{E}_\text{ext}|^2+|c\vec{B}_\text{ext}|^2\right)$ is a constant energy offset that will be ignored. For alkali Rydberg atoms the internal magnetic field is negligible: the valence electron moves slowly due to a weak centripetal force from the Coulomb potential. We may therefore drop $\vec{B}_\text{int}$. We may also decompose the electric fields as $\vec{E}=\vec{E}^\parallel+\vec{E}^\perp$ where $\vec{\nabla}\cdot\vec{E}^\perp=0$ and $\vec{\nabla}\times\vec{E}^\parallel=0$ and identity that the external field is transverse by Gauss' law, because the sources that produce it are far from the atom. The Hamiltonian becomes
\begin{align}
    H_\text{PZW}&=\frac{(\vec{p}-\vec{\Xi}_\text{ext})^2}{2m_e}+\epsilon_0\int d^3\vec{r}\,\vec{E}_\text{int}^\perp\cdot\vec{E}_\text{ext}^\perp-\vec{\mu}_S\cdot \vec{B}_\text{ext}(\vec{r}_e)\nonumber\\&+\frac{\epsilon_0}{2}\int d^3\vec{r}\,|\vec{E}_\text{int}^\perp|^2+\frac{\epsilon_0}{2}\int d^3\vec{r}\,|\vec{E}_\text{int}^\parallel|^2
\end{align}
with the second-to-last integral being a transverse self energy that is a constant energy offset that will be ignored, and the final integral is the Coulomb potential $V_\text{Coul}$. 

A simple choice for $\vec{E}_\text{int}$ that satisfies Maxwell's equation is to write it as a path integral over the straight line $\vec{s}_\text{SL}(\lambda)=\vec{r}_c+\lambda\vec
r_{ec}$ between the core and the valence electron. The internal fields become
\begin{align}
    \vec{E}_\text{int}(\vec{r})&=-\frac{\vec{d}}{\epsilon_0}\int_0^1 d\lambda \,\delta(\vec{r}-\vec{s}_\text{SL}(\lambda))\\
    c\vec{B}_\text{int}(\vec{r})&=\frac{2\vec{\mu}_\text{orb}}{\epsilon_0 c}\left(\int_0^1 d\lambda\,\lambda\delta(\vec{r}-\vec{s}_\text{SL}(\lambda))+\vec{\mu}_S\delta(\vec{r}-\vec{r}_e)\right)
\end{align}
where $\vec{d}=-e\vec{r}_{ec}$ and $\vec{\mu}_\text{orb}=\frac{\vec{d}\times \vec{p}_{ec}}{2m_e}$ are the electric and orbital magnetic dipole moments of the atom. Defining new external fields that are weighted averages along the straight line, $\langle \vec{E}_\text{ext}\rangle=\int_0^1 d\lambda\, \vec{E}_\text{ext}(\vec{s}_\text{SL}(\lambda))$ and $\langle \vec{B}_\text{ext}\rangle=2\int_0^1 d\lambda\, \lambda\vec{B}_\text{ext}(\vec{s}_\text{SL}(\lambda))$, the PZW Hamiltonian simplifies to an internal term and four atom-field interaction terms:
\begin{align}
    H_\text{PZW}=&H_\text{int}+V_\text{PZW}\\
    H_\text{int}=&\frac{|\vec{p}|^2}{2m_e}+V_\text{Coul}\\
    V_\text{PZW}=&-\vec{\mu}_\text{orb}\cdot\langle\vec{B}_\text{ext}\rangle-\vec{\mu}_S\cdot \vec{B}_\text{ext}(\vec{r}_e)\nonumber\\&+\frac{|\vec{d}\times\langle\vec{B}_\text{ext}\rangle|^2}{8m_e}-\vec{d}\cdot\langle\vec{E}_\text{ext}\rangle
\end{align}
These terms are the orbital magnetic-dipole interaction, the spin magnetic-dipole interaction, the diamagnetic interaction, and the electric-dipole interaction. Remarkably, the Hamiltonian is in the same form as the dipole approximation, and it reduces to that approximation if the external fields are uniform across the atom. This derivation omitted relativistic effects, whose first-order correction produces a spin-orbit coupling term in $H_\text{int}$.
\section{Floquet Photon Expansion}
\label{appendix:FloquetPhoton}
The space of functions with period $T=\frac{2\pi}{\omega}$, $\mathcal{T}$, can be spanned by a time basis $\ket{t}$ or a photon basis $\ket{n}=\frac{1}{T}\int_0^T dt e^{in\omega t}\ket{t}$. The inner product of both bases is $\langle t' \mid t\rangle=T\delta(t'-t)$ and $\langle n'|n\rangle=\delta_{n'n}$, respectively \cite{Rodriguez_2018_TimeBasis}. The Hilbert space may be expanded to Sambe space, defined as $\mathcal{S}\equiv\mathcal{H}\otimes \mathcal{T}$ \cite{Sambe_1973_Floquet}. The time-independent Hamiltonian $H_\mathcal{S}$ can be expressed with $\mathcal{T}$ in either the time or photon basis as follows:
\begin{align}
    H_\mathcal{S}&=\int_0^T \frac{dt}{T}(H(t)\otimes\ket{t}\bra{t})-\left(\mathbb{I}_\mathcal{H}\otimes i\hbar\frac{\partial}{\partial t}\right)\\
    &=\sum_{n,k}\left(H^{k}\otimes\ket{n+k}\bra{n}+\mathbb{I}_\mathcal{H}\otimes n\hbar\omega\ket{n}\bra{n}\right)
\end{align}
Here $H^k$ is the Fourier component of the time-dependent Hamiltonian $H(t)$:
\begin{align}
    H^k&=\int^T_0 \frac{dt}{T}H(t)e^{-ik\omega t}
\end{align}
We proceed to use the photon basis with the interpretation that operators $H^k$ can add $k$ photons at an energy cost of $k\hbar\omega$. Specifically, the AC fields may be written in the Sambe space as
\begin{align} 
\vec{E}_{\text{ac}}(\vec{r})
&= \sum_{n} \left(\frac{\vec{\mathcal{E}}_{\text{ac}}(\vec{r})}{2}\otimes\ket{n-1}\bra{n}+\frac{\vec{\mathcal{E}}^*_{\text{ac}}(\vec{r})}{2}\otimes\ket{n+1}\bra{n}\right) 
\end{align}
with the interpretation that $\vec{\mathcal{E}}_\text{ac}$ and $\vec{\mathcal{E}}_\text{ac}^*$ drive photon absorption and emission, respectively. 

Performing second order perturbation theory with this Hamiltonian, as described in Appendix \ref{appendix:PerturbationTheory}, consists of processes where the system leaves and returns to the same initial fine structure level $\ket{nLSJ}\otimes\ket{n}$. Thus, $\vec{\mathcal{E}}_\text{ac}$ must always be paired with $\vec{\mathcal{E}}^*_\text{ac}$ so that the sum of photon emissions and absorptions is zero. The virtual state energy contains $\hbar\omega$ excess (reduced) energy upon virtual emission (absorption), which explains the factors of $\hbar\omega$ in the resolvent of equation \ref{eq:resolvent}.
\section{Finite Lifetime}
\label{appendix:FiniteLifetime}

The weak coupling of an atom to its environment without memory or interference is well approximated by the Lindblad master equation:
\begin{align}
    \frac{d\rho}{dt}=-\frac{i}{\hbar}[H,\rho]-\frac{1}{2}\{\Gamma,\rho\}+\mathcal{J}
\end{align}
where $\rho$ is the density operator of the atom, $\Gamma\equiv \sum_k L^\dagger_k L_k$ is the angular decay rate operator,  $\mathcal{J}\equiv \sum_k L_k \rho L_k^\dagger$ is the jump super-operator, and $L_k$ is the Lindblad jump operator. The final term serves to repopulate states after a decay event, which we approximate away. We assume the states in $\mathcal{P}$ have no decay, and that population in the states of $\mathcal{Q}$ is negligible, such we can ignore where that population decays. The master equation reduces to a Schrödinger equation whose effective Hamiltonian is not Hermitian:
\begin{align}
    H_\text{eff}=H-i\hbar\Gamma/2
\end{align}
The consequence for polarizability theory is that the resolvent (\ref{eq:resolvent}) is redefined as 
\begin{align}
    R^{(0)}_k&=-\frac{Q}{\hbar\Delta_k-i\hbar\Gamma/2}
\end{align}
whose Neumann expansion (\ref{eq:ResolventNeumannExpansion}) is redefined as
\begin{align}
    R^{(0)}_k=\sum_{m=1}^\infty (-1)^m \frac{(\hbar\Delta_0-i\hbar\Gamma/2)^{m-1}}{(k\hbar\omega)^m}\label{eq:resolventNeumannExpansionGamma}
\end{align}
Crucially, observe that the real part of the resolvent is unaffected by $\Gamma$ for $\omega^{-1}$ and $\omega^{-2}$ dependence, such that the cancellation of the real $\omega^{-1}$ rank-1 (\ref{eq:vecCancellation}) and real $\omega^{-2}$ rank-2 terms (\ref{eq:tenCancellation}) remains valid. 

The expression for the resonance term (\ref{eq:GammaNew}) must be modified accordingly. We replace $\omega_q$ with $\omega_q-i\Gamma_q/2$ and subtract the analytically derived `zero' terms to obtain
\begin{align}
    \gamma^{(0)}_q(\omega)&=\frac{\omega_q-i\Gamma_q/2}{\hbar(\omega^2-(\omega_q-i\Gamma_q/2)^2)}\nonumber\\
    &=\frac{\omega_q \chi}{\hbar \zeta} - i \left( \frac{\Gamma_q \sigma}{2\hbar \zeta} \right)\label{eq:GammaGamma0}\\
    \gamma^{(1)}_q(\omega)&=-\frac{\omega}{\hbar(\omega^2-(\omega_q-i\Gamma_q/2)^2)}-\left(-\frac{1}{\hbar\omega}\right)\nonumber\\
    &= \left[ \frac{1}{\hbar\omega} - \frac{\omega \rho}{\hbar \zeta} \right] + i \left( \frac{\omega \omega_q \Gamma_q}{\hbar \zeta} \right)\label{eq:GammaGamma1}\\
    \gamma^{(2)}_q(\omega)&=\frac{\omega_q-i\Gamma_q/2}{\hbar(\omega^2-(\omega_q-i\Gamma_q/2)^2)}-\left(\frac{\omega_q}{\hbar\omega^2}\right)\nonumber\\
    &=\left[ \frac{\omega_q \chi}{\hbar \zeta} - \frac{\omega_q}{\hbar\omega^2} \right] - i \left( \frac{\Gamma_q \sigma}{2\hbar \zeta} \right)\label{eq:GammaGamma2}
\end{align}
Due to the notational length required to separate the resonance terms into their real and imaginary parts, we introduce the intermediate variables $\chi = \omega^2 - \omega_q^2 - \frac{\Gamma_q^2}{4}$, $\rho = \omega^2 - \omega_q^2 + \frac{\Gamma_q^2}{4}$, $\sigma = \omega^2 + \omega_q^2 + \frac{\Gamma_q^2}{4}$, and $\zeta = \rho^2 + \omega_q^2\Gamma_q^2$. The expressions (\ref{eq:GammaGamma0}-\ref{eq:GammaGamma2}) are used to calculate the real and imaginary polarizabilities in Fig. \ref{fig:nearDetuned}.

The near-detuned regime is defined by single level of $\mathcal{Q}$, $\ket{q'}=\ket{n'L'SJ'}$, being near-resonant with the level that defines $\mathcal{P}$. The detuning $\delta_{q'}=\omega-|\omega_{q'}|$ is much smaller than the optical frequency, yet larger than the linewidth of the admixed level: $\Gamma_{q'}\ll |\delta_{q'}|\ll\omega$. The dominant resonance term simplifies significantly to
\begin{align}
    \gamma^{(\kappa)}_{q'}(\omega)=[\text{sgn}(\omega_{q'})]^{\kappa+1}\left[\frac{1}{\hbar\delta_{q'}}-i\frac{\Gamma_{q'}/2}{\hbar\delta^2_{q'}}\right]\label{eq:GammaNearDetunedAppendix}
\end{align}
where the sign function is positive(negative) for a virtual state of higher(lower) energy. Equation (\ref{eq:GammaNearDetuned}Appendix) reproduces the intuition from a two-state system that perturbations scale as $\delta^{-1}$ and scattering as $\delta^{-2}$.
\section{Cartesian and Spherical Tensors}
\label{appendix:CartesianSpherical}
We provide a definition for the irreducible projection of a dyadic of the form $\left[\vec{A}\otimes\vec{B}\right]^{(\kappa)}$.  The spherical basis consists of the eigenstates of the $\hat{L}_Z$ operator in a space of unit angular momentum with the Condon-Shortley phase convention. The covariant spherical basis is labeled in the form $\ket{\kappa,q}\equiv e_{\kappa,q}$ as
\begin{align}
    \begin{matrix}
        \ket{1,-1}\\\ket{1,0}\\\ket{1,1}
    \end{matrix}&=\begin{pmatrix}
        \frac{1}{\sqrt{2}}&-\frac{i}{\sqrt{2}}&0\\0&0&1\\-\frac{1}{\sqrt{2}}&-\frac{i}{\sqrt{2}}&0
    \end{pmatrix}\begin{matrix}
        \ket{x}\\\ket{y}\\\ket{z}
    \end{matrix}
\end{align}
where $\ket{i}\equiv e_i$ is the Cartesian covariant basis. We also define a spherical basis for the dyadic as a spherical tensor product [\cite{Zare1988}]:
\begin{align}
    \ket{(1,1)\kappa q}&=\sum_{q_1 q_2}C_{\kappa q}^{1 q_1 1 q_2}\ket{1 q_1}\otimes\ket{1 q_2}
\end{align}
For reference we evaluate the $\kappa=\{0,1\}$ states:
\begin{align}
    \ket{(11)0,0}&=-\frac{1}{\sqrt{3}}\sum_{i=1}^3\ket{ii}=-\frac{1}{\sqrt{3}}\mathbb{I}\\
    \ket{(11)1,0}&=\frac{i}{\sqrt{2}}\left(\ket{xy}-\ket{yx}\right)\\
    \ket{(11)1,\pm 1}&=\frac{i}{\sqrt{2}}\left(\frac{i(\ket{xz}-\ket{zx})\mp(\ket{yz}-\ket{
    zy})}{\sqrt{2}}\right)\label{eq:11kq}
\end{align}
where we use the shorthand notation $\ket{ij}\equiv \ket{i}\otimes\ket{j}$. 

We are now able to define the irreducible projection of the dyadic as the projection onto the spherical basis of rank $\kappa$:
\begin{align}
    \left[\vec{A}\otimes\vec{B}\right]^{(\kappa)}&=
\left(\vec{A}\otimes\vec{B}\right)P_\kappa\\&=\sum_{q=-\kappa}^\kappa \vec{A}\otimes\vec{B}\ket{(11)\kappa, q}\bra{(11)\kappa,q}
\end{align}
The choice of placing the projector $P_\kappa\equiv\sum_q \ket{(11)\kappa,q}\bra{(11)\kappa,q}$ on the right of the dyadic ensures the convention that a spherical tensor has covariant components and a contravariant basis \cite{Man2013_TensorCartSphr}. 

We further simplify the dyadic into its Cartesian components $\vec{A}\otimes\vec{B}=\sum_{ij}A_iB_j\bra{ij}$ to obtain
\begin{align}
    \left[\vec{A}\otimes\vec{B}\right]^{(\kappa)}&=\sum_{ijq}A_iB_j\bra{ij}(11)\kappa,q\rangle\bra{(11)\kappa,q}
\end{align}
We use (\ref{eq:11kq}) to evaluate $\bra{ij}(11)\kappa,q\rangle$ for $\kappa=\{0,1\}$. We obtain
\begin{align}
    \left[\vec{A}\otimes\vec{B}\right]^{(0)}&=-\frac{1}{\sqrt{3}}\vec{A}\cdot \vec{B}\bra{(11)0,0}\label{eq:rank0Dyadic}\\
    \left[\vec{A}\otimes\vec{B}\right]^{(1)}&=\sum_q \frac{i}{\sqrt{2}}\left[\vec{A}\times\vec
    B\right]^1_q\bra{(11)1,q}\label{eq:rank1Dyadic}
\end{align}
Both $\bra{(11)0,0}$ and $\bra{(11)1,q}$ are composed of the dyadic basis, whereas $\bra{00}\equiv1$ and $\bra{1,q}\equiv (-1)^q \ket{1,-q}$ are composed of the scalar, and vector bases, respectively. The beauty of spherical tensors is that the terms are isomorphic, allowing us to simplify the basis in (\ref{eq:rank0Dyadic}) and (\ref{eq:rank1Dyadic}) without any functional change to the object's behavior. Therefore, we may write the projection of the dyadic as a scalar for $\kappa=0$ and as a vector for $\kappa=1$:
\begin{align}
    \left[\vec{A}\otimes\vec{B}\right]^{(\kappa)}&\cong-\frac{1}{\sqrt{3}}\left(\vec{A}\cdot\vec{B}\right)\delta_{\kappa,0}+\frac{i}{\sqrt{2}}\left(\vec{A}\times\vec{B}\right)\delta_{\kappa,1}\nonumber\\&+\left[\vec{A}\otimes\vec{B}\right]^{(2)}\delta_{\kappa,2}
\end{align}
We note that the Kronecker delta implies the right hand side is a piecewise function, and it is not the sum of a scalar, vector, and tensor. 
\section{The Liouvillian}
\label{appendix:Liouvillian}

In section \ref{sec:NotBeyondDipole} we expand the polarizability operator in powers of $\omega^{-1}$ and find each term in the expansion to be proportional to the operator $P\vec{d}\otimes (H_0-E_0)^m\vec{d}P$. Here we prove that such terms may be redefined in terms of the Liouvillian super-operator, defined for a closed system by $\mathcal{L}(\hat{O})=\frac{i}{\hbar}[H_0,\hat{O}]$. The redefinition is possible by converting $E_0$ into an operator using $E_0P=H_0P$ and $PH_0=PE_0$. 

Both a rightward and leftward evaluation of $H_0-E_0$ produce the Liouvillian:
\begin{align}
    (H_0-E_0)\vec{d}P&=\left(H_0\vec{d}-\vec{d}H_0\right)P\nonumber\\
    &=[H_0,\vec{d}]P\nonumber\\
    &=(-i\hbar)\mathcal{L}(\vec{d})P\\
    P\vec{d}(H_0-E_0)&=P(\vec{d}H_0-H_0\vec{d})\nonumber\\
    &=-P[H_0,\vec{d}]\nonumber\\
    &=(-1)^1P(-i\hbar)\mathcal{L}(\vec{d})
\end{align}
We assume the pattern holds for $m-1$ repeated rightward and leftward evaluations:
\begin{align}
    (H_0-E_0)^{m-1}\vec{d}P &= (-i\hbar)^{m-1} \mathcal{L}^{m-1}(\vec{d})P\\
    P\vec{d}(H_0-E_0)^{m-1}&=(-1)^{m-1}P(-i\hbar)^{m-1}  \mathcal{L}^{m-1}(\vec{d})
\end{align}
Applying an additional operation proves the rightward relationship to be true by induction.
\begin{align}
    (H_0-E_0)^{m}\vec{d}P &= (H_0-E_0)(-i\hbar)^{m-1} \mathcal{L}^{m-1}(\vec{d})P\nonumber\\
    &=(-i\hbar)^{m-1} (H_0\mathcal{L}^{m-1}(\vec{d})-\mathcal{L}^{m-1}(\vec{d})H_0)P\nonumber\\
    &=(-i\hbar)^{m-1} [H_0,\mathcal{L}^{m-1}(\vec{d})]P\nonumber\\
    &=(-i\hbar)^{m}\mathcal{L}^{m}(\vec{d})P
\end{align}
Similarly the leftward relationship is proven by induction to be 
\begin{align}
    P\vec{d}(H_0-E_0)^m&=P(i\hbar)^{m}  \mathcal{L}^{m}(\vec{d})
\end{align}
The operator we seek to evaluate, $P\vec{d}\otimes (H_0-E_0)^m\vec{d}P$, may therefore be rewritten as $a$ leftward operations and $b$ rightward operations where $a+b=m$.
\begin{align}
    P\vec{d}\otimes (H_0-E_0)^m\vec{d}P&=P\vec{d}(H_0-E_0)^a\otimes(H_0-E_0)^b\vec{d}P\nonumber\\
    &=(-1)^b(i\hbar)^mP\mathcal{L}^a(\vec{d})\otimes\mathcal{L}^b(\vec{d})P
\end{align}
One could have equivalently performed $b$ leftward operations and $a$ rightward operations, resulting in the following exchange identity.
\begin{align}
   \mathcal{L}^a(\vec{d})\otimes\mathcal{L}^b(\vec{d})&=(-1)^m \mathcal{L}^b(\vec{d})\otimes\mathcal{L}^a(\vec{d})
\end{align}
Any outer product may be written as an average of its symmetric and anti-symmetric parts, which may be simplified using the exchange identity:
\begin{align}
     &\mathcal{L}^a(\vec{d})\otimes\mathcal{L}^b(\vec{d})=\frac{\left\{\mathcal{L}^a(\vec{d}),\mathcal{L}^b(\vec{d})\right\}_\otimes+\left[\mathcal{L}^a(\vec{d}),\mathcal{L}^b(\vec{d})\right]_\otimes}{2}\nonumber\\
     &=\frac{\delta_{m,\text{even}}}{2}\left\{\mathcal{L}^a(\vec{d}),\mathcal{L}^b(\vec{d})\right\}_\otimes+\frac{\delta_{m,\text{odd}}}{2}\left[\mathcal{L}^a(\vec{d}),\mathcal{L}^b(\vec{d})\right]_\otimes
\end{align}
We thus obtain equation \ref{eq:LiouvillianIdentity}.
% \begin{align}
%     P\vec{d}\otimes (H_0-E_0)^m\vec{d}P
%     &=\frac{(-1)^b}{2}(i\hbar)^mP\left(\delta_{m,\text{even}}\left\{\mathcal{L}^a(\vec{d}),\mathcal{L}^b(\vec{d})\right\}_\otimes+\delta_{m,\text{odd}}\left[\mathcal{L}^a(\vec{d}),\mathcal{L}^b(\vec{d})\right]_\otimes\right)P
% \end{align}

\section{The Van Vleck High Frequency Expansion}
\label{appendix:VV}
An atom in a monochromatic optical field is described by the time-dependent Schrödinger equation
\begin{align}
    i\frac{d}{dt}\ket{\psi(t)}&=H(t)\ket{\psi(t)}\label{eq:TDSE}
\end{align}
with a time-periodic Hamiltonian $H(t)=H(t+T)$ where $\omega=2\pi/T$. The Hamiltonian has Fourier components $H_k$ such that $ H(t)=\sum_{k} H_k e^{i k\omega t}$ with the property $H_{-k}=H^\dagger_k$.  Each state undergoes motion at two time scales: fast micromotion, and slow macromotion. The micromotion must be periodic with respect to $T$, so we can write a state as
\begin{align}
    \ket{\psi(t)}&=e^{-iK(t)}\ket{\phi(t)}\label{eq:Macromotion}
\end{align}
where $\ket{\phi(t)}$ is affected by macromotion that evolves under a time independent Hamiltonian $H_\text{VV}$, and $K(t)$ is periodic with no time average, meaning it returns the system to the state $\ket{\phi(t)}$ after every period \cite{VanVleck_1932_TheoryBook}. Inserting \ref{eq:Macromotion} into \ref{eq:TDSE} produces
\begin{align}
    i\frac{d}{dt}\ket{\phi(t)}&=H_\text{VV}\ket{\phi(t)}\\
       H_\text{VV}&\equiv e^{iK(t)}H(t)e^{-iK(t)}-ie^{iK(t)}\left(\frac{d}{dt}e^{-iK(t)}\right)
\end{align}
Although $H_\text{VV}$ appears time dependent, we solve for $K(t)$ precisely so that $\ket{\phi(t)}$ evolves under a time-independent Hamiltonian. The time derivative of the micromotion unitary evaluates to
\begin{align}
    \partial_t e^{K(t)}&=\int_0^1 e^{\lambda K(t)}\dot{K}(t)e^{(1-\lambda)K(t)}d\lambda
\end{align}
such that
\begin{align}
    e^{-K(t)}\partial_t e^{K(t)}&=\int_0^1 e^{-\lambda K(t)}\dot{K}(t)e^{\lambda K(t)}d\lambda
\end{align}
We also use the Baker-Campbell-Hausdorff lemma
\begin{align}
    e^A B e^{-A}&=\sum_{k=0}^\infty \frac{1}{k!}\text{ad}^k_A(B)
\end{align}
defined in terms of $\text{ad}_A(B)=[A,B]$. The Hamiltonian becomes
\begin{align}
    H_\text{VV}&=\sum_{k=0}^\infty h_{k}\\
    h_{k}&\equiv \frac{i^k}{k!}\left(\text{ad}^k_{K(t)}(H(t))-\frac{1}{k+1}\text{ad}^k_{K(t)}(\dot{K}(t))\right)
\end{align}
The first three terms are
\begin{align}
 h_{0}&=H(t)-\dot{K}(t)\\
   h_{1}&=i[K(t),H(t)]-\frac{i}{2}[K(t),\dot{K}(t)]\\
    h_{2}&=-\frac{1}{2}[K(t),[K(t),H(t)]]+\frac{1}{6}[K(t),[K(t),\dot{K}(t)]]
\end{align}
The terms $H_\text{VV}$, $H(t)$, and $K(t)$ may be expanded in powers of $1/\omega$, such as
\begin{align}
    H(t)&=\sum_{w=0}^\infty H^{(w)}(t)=\sum_{w,n=0}^\infty H^{(w)}_ne^{in\omega t}
\end{align}
Other derivations \cite{Eckardt_2015_HighFrequencyFloquet} omit the expansion of $H(t)$, which we must retain due to the explicit $\omega^{-1}$ dependence of magnetic terms in the Hamiltonian. We begin an iterative process starting with an initial guess that there is no micromotion by setting $\omega\rightarrow\infty$. Equivalently, $K^{(0)}=0$. The commutators in $k\neq 0$ collapse and we obtain
\begin{align}
    H^{(0)}_\text{VV}&= \langle h^{(0)}_{0}\rangle=H^{(0)}_0
\end{align}
The term $\dot{K}^{(1)}(t)$ must absorb the time dependence of $h^{(0)}_0$, such that
\begin{align}
    \dot{K}^{(1)}(t)
    &=\sum_{n\neq 0}H^{(0)}_ne^{in\omega t}
\end{align}
resulting in
\begin{align}
    K^{(1)}(t)&=\sum_{n\neq 0}H^{(0)}_n\int e^{in\omega t}dt\nonumber\\
    &=-i\sum_{n\neq 0}H^{(0)}_n\frac{e^{i n\omega t}}{n\omega}
\end{align}
We set the integration constant to zero because the micromotion has no mean. Iterating this process for $H^{(1)}_\text{VV}$ and $H^{(2)}_\text{VV}$ produces the following Van Vleck Hamiltonian:
\begin{align}
    H_\text{VV}&=H^{(0)}_0+H^{(1)}_0+\sum_{n\neq 0}\frac{[H^{(0)}_n,H^{(0)}_{-n}]}{2n\hbar\omega}+\sum_{n\neq 0}\frac{[H^{(0)}_n,H^{(1)}_{-n}]}{n\hbar\omega}\nonumber\\&+H^{(2)}_0-\sum_{n\neq 0}\frac{1}{2n^2\hbar^2\omega^2}[H^{(0)}_n,[H^{(0)}_{-n},H^{(0)}_0]]\nonumber\\&+\sum_{\substack{n,m\neq 0\\n+m\neq 0}}\frac{1}{3nm\hbar^2\omega^2}[H^{(0)}_n,[H^{(0)}_m,H^{(0)}_{-n-m}]]+\mathcal{O}(\omega^{-3})
\end{align}

\end{document}